\documentclass[11pt,a4paper]{article}
\pdfoutput=1

\usepackage{jheppub}
\usepackage[utf8]{inputenc}
\usepackage[T1]{fontenc}
\usepackage{afterpage}
\usepackage{rotating}

\newcommand{\eVq}{\ensuremath{\text{eV}^2}}
\newcommand{\Dmq}{\Delta m^2}
\newcommand{\Eps}{\varepsilon}
\newcommand{\Epx}{\mathcal{E}}
\newcommand{\dd}{\mathrm{d}}
\newcommand{\Mmed}{M_\text{med}}

\renewcommand{\Re}{\mathop{\mathrm{Re}}}
\renewcommand{\Im}{\mathop{\mathrm{Im}}}

\DeclareMathOperator{\Tr}{Tr}
\DeclareMathOperator{\diag}{diag}

\makeatletter
\DeclareRobustCommand\recite[1]{\begingroup\@fileswfalse\cite{#1}\endgroup}
\makeatother

\graphicspath{{Figures/}}

\allowdisplaybreaks

\title{Global constraints on non-standard neutrino interactions with
  quarks and electrons}

\author[a]{Pilar Coloma,}
\affiliation[a]{Instituto de F\'isica Te\'orica (IFT-CFTMAT),
  CSIC-UAM, Calle de Nicol\'as Cabrera 13--15, Campus de Cantoblanco,
  E-28049 Madrid, Spain}
\emailAdd{pilar.coloma@ift.csic.es}

\author[b,c,d]{M.~C.~Gonzalez-Garcia,}
\affiliation[b]{C.N.~Yang Institute for Theoretical Physics, Stony
  Brook University, Stony Brook, NY 11794-3840, USA}
\affiliation[c]{Departament de F\'isica Qu\`antica i Astrof\'isica and
  Institut de Ci\`encies del Cosmos, Universitat de Barcelona,
  Diagonal 647, E-08028 Barcelona, Spain}
\affiliation[d]{Instituci\'o Catalana de Recerca i Estudis Avançats
  (ICREA), Pg.\ Lluis Companys 23, E-08010 Barcelona, Spain}
\emailAdd{maria.gonzalez-garcia@stonybrook.edu}

\author[a]{Michele Maltoni,}
\emailAdd{michele.maltoni@csic.es}

\author[c]{Jo\~ao Paulo Pinheiro,}
\emailAdd{joaopaulo.pinheiro@fqa.ub.edu}

\author[e]{Salvador Urrea}
\emailAdd{salvador.urrea@ific.uv.es}
\affiliation[e]{Instituto de F\'isica Corpuscular (IFIC), CSIC-UV,
  Edificio Institutos de Investigaci\'on, Calle Catedr\'atico Jos\'e
  Beltr\'an 2, E-46980 Paterna, Spain}

\abstract{We derive new constraints on effective four-fermion neutrino
  non-standard interactions with both quarks and electrons.  This is
  done through the global analysis of neutrino oscillation data and
  measurements of coherent elastic neutrino-nucleus scattering
  (CE$\nu$NS) obtained with different nuclei.  In doing so, we include
  not only the effects of new physics on neutrino propagation but also
  on the detection cross section in neutrino experiments which are
  sensitive to the new physics.  We consider both vector and
  axial-vector neutral-current neutrino interactions and, for each
  case, we include simultaneously all allowed effective operators in
  flavour space.  To this end, we use the most general parametrization
  for their Wilson coefficients under the assumption that their
  neutrino flavour structure is independent of the charged fermion
  participating in the interaction.  The status of the LMA-D solution
  is assessed for the first time in the case of new interactions
  taking place simultaneously with up quarks, down quarks, and
  electrons.  One of the main results of our work are the presently
  allowed regions for the effective combinations of non-standard
  neutrino couplings, relevant for long-baseline and atmospheric
  neutrino oscillation experiments.}

\preprint{IFT-UAM/CSIC-23-47, IFIC/23-15, FTUV-23-0427.3710, YITP-SB-2023-05}

\keywords{neutrinos, non-standard interactions, coherent elastic
  neutrino-nucleus scattering, oscillations}

\begin{document}

\maketitle

\section{Introduction}

The minimal global description of the bulk of data gathered in
experiments detecting the interactions of solar and atmospheric
neutrinos, and of neutrinos produced in nuclear reactors and in
particle accelerators, requires three neutrino states with distinct
masses which are non-trivial admixtures of the three flavour neutrino
states of the Standard Model (SM).  This implies that lepton flavour
oscillates with a wavelength which depends on distance and
energy~\cite{Pontecorvo:1967fh, Gribov:1968kq} as required to explain
the data, see Ref.~\cite{GonzalezGarcia:2007ib} for an overview.  This
is one of the most direct indications of physics beyond the Standard
Model (BSM).

Given the excellent ability of the SM in explaining the interactions
of fundamental particles through the electromagnetic, weak, and strong
interactions, it is reasonable to assume that the SM is an effective
theory and that new physics (NP) effects are suppressed at low
energies.  Generic BSM physics can then be introduced in a
model-independent manner through a tower of effective operators,
suppressed by the heavy NP scale.  Interestingly enough, neutrino
masses arise in this framework from the unique dimension-five operator
consistent with the SM gauge symmetry and particle contents, and
therefore can be argued to be the first signal of BSM physics.  At
next order in the operator expansion we find dimension-six
four-fermion operators.  Those involving neutrino fields would affect
the production, propagation, and detection of neutrinos --~the
so-called Non-Standard neutrino Interactions (NSI).  In this work we
focus on operators leading to purely vector or axial-vector
interactions (for recent works including additional operators with
different Lorentz structures see for example
Refs.~\cite{Falkowski:2021bkq, Breso-Pla:2023tnz, Falkowski:2019kfn,
  Falkowski:2019xoe}).  Generically these can be classified in
charged-current (CC) NSI
\begin{equation}
  \label{eq:nsi-cc}
  \mathcal{L}_\text{NSI,CC} = -2\sqrt{2}\, G_F
  \sum_{f,f^\prime,\alpha,\beta}
  \Eps_{\alpha\beta}^{ff^\prime,P}
  (\bar\ell_\alpha \gamma_\mu P_L \nu_\beta)
  (\bar f \gamma^\mu P f^\prime)
  + \text{h.c.}
\end{equation}
and neutral-current (NC) NSI
\begin{equation}
  \label{eq:nsi-nc}
  \mathcal{L}_\text{NSI,NC} = -2\sqrt{2}\, G_F
  \sum_{f,P,\alpha,\beta} \Eps_{\alpha\beta}^{f,P}
  (\bar\nu_\alpha\gamma^\mu P_L\nu_\beta)
  (\bar f\gamma_\mu P f) \,.
\end{equation}
In Eqs.~\eqref{eq:nsi-cc} and~\eqref{eq:nsi-nc}, $f$ and $f^\prime$
refer to SM charged fermions, $\ell$ denotes a SM charged lepton and
$P$ can be either a left-handed or a right-handed projection operator
($P_L$ or $P_R$, respectively).  Moreover the normalization of the
couplings is deliberately chosen to match that of the weak currents in
the SM, so the values of $\Eps_{\alpha\beta}^{f,P}$ indicate the
strength of the new interaction with respect to the Fermi constant,
$G_F$.
The corresponding vector and axial-vector combinations of NSI
coefficients are defined as:
\begin{equation}
  \label{eq:vect-axial}
  \Eps_{\alpha\beta}^{f,V} \equiv
  \Eps_{\alpha\beta}^{f,L} + \Eps_{\alpha\beta}^{f,R}
  \quad\text{and}\quad
  \Eps_{\alpha\beta}^{f,A} \equiv
  \Eps_{\alpha\beta}^{f,L} - \Eps_{\alpha\beta}^{f,R} \,.
\end{equation}

Precise measurements of meson and muon decays place severe constraints
on the possible strength of CC NSI (see for example
Refs.~\cite{Davidson:2003ha, Biggio:2009nt, Biggio:2009kv,
  Falkowski:2021bkq}).  Conversely, NC NSI are much more difficult to
probe directly, given the intrinsic difficulties associated to
neutrino detection via neutral currents.  A priori, the requirement of
gauge invariance would generate similar operators in the charged
lepton sector, in severe conflict with experimental
observables~\cite{Gavela:2008ra, Antusch:2008tz}.  However, such
bounds may be alleviated (or evaded) in NP models in which NC NSI are
generated by exchange of neutral mediators with masses well below the
EW scale (for a recent review on viable NSI models see, \textit{e.g.},
Ref.~\cite{Dev:2019anc}).  It is in these scenarios that one can
envision observable effects in present and future neutrino
experiments.
Several models have been proposed, involving new gauge symmetries and
light degrees of freedom, that would give rise to relatively large NC
NSI.  These include, for example, models where the NSI are generated
from the $Z'$ boson associated to a new $U(1)'$
symmetry~\cite{Babu:2017olk, Farzan:2015doa, Farzan:2016wym,
  Farzan:2015hkd, Greljo:2022dwn, Heeck:2018nzc, Farzan:2019xor,
  Bernal:2022qba}, radiative neutrino mass models involving new
scalars~\cite{Babu:2019mfe}, or models with
leptoquarks~\cite{Wise:2014oea, Greljo:2021npi, Babu:2019mfe}.  Many
of these extensions involve a gauge symmetry based on a combination of
baryon and lepton quantum numbers, and would therefore induce equal
NSI for up and down quarks.  However, exceptions to this rule arise,
for example, in models with leptoquarks where NSI may be only
generated for down-quarks~\cite{Wise:2014oea, Babu:2019mfe}.
Similarly it should be stressed that, while many of these models
typically lead to diagonal NSI in lepton flavor space, this is not
always the case and, depending on the particular extension, sizable
off-diagonal NSI may also be obtained (see, \textit{e.g.},
Refs.~\cite{Farzan:2015hkd, Farzan:2019xor, Farzan:2016wym}).  In
order to derive constraints to such a wide landscape of models, the
use of the effective operator approach advocated above is extremely
useful.  For bounds from oscillation data on $U(1)'$ models with
diagonal couplings in flavor space, see Ref.~\cite{Coloma:2020gfv}.

In fact, some of the best model-independent bounds on NC NSI are
obtained from global fits to oscillation data.  These are affected by
vector couplings involving fermions present in matter,
$\Eps_{\alpha\beta}^{f,V}$ with $f \in \{u,d,e \}$, since they modify
the effective matter potential~\cite{Wolfenstein:1977ue,
  Mikheev:1986gs} felt by neutrinos as they propagate in a medium.
In Ref.~\cite{Gonzalez-Garcia:2013usa} such global analysis was
performed in the context of vector NC NSI with either up or down
quarks.  In Ref.~\cite{Esteban:2018ppq} the study was extended to
account for the possibility vector NC NSI with up and down quarks
simultaneously, under the restriction that the neutrino flavour
structure of the NSI interactions is independent of the quark type.
However, NSI with electrons were not considered.

One important effect of the presence of NSI~\cite{Wolfenstein:1977ue,
  Valle:1987gv, Guzzo:1991hi} affecting the neutrino propagation in
oscillation experiments is the appearance of a degeneracy leading to a
qualitative change of the lepton mixing pattern.  This was first
observed in the context of solar neutrinos, for which the established
standard Mikheev-Smirnov-Wolfenstein (MSW)
solution~\cite{Wolfenstein:1977ue, Mikheev:1986gs} requires a mixing
angle $\theta_{12}$ in the first octant, while with suitable NSI the
data could be described by a mixing angle $\theta_{12}$ in the second
octant, the so-called LMA-Dark (LMA-D)~\cite{Miranda:2004nb} solution.
The origin of the LMA-D solution is a degeneracy in the oscillation
probabilities due to a symmetry of the Hamiltonian describing neutrino
evolution in the presence of NSI~\cite{GonzalezGarcia:2011my,
  Gonzalez-Garcia:2013usa, Bakhti:2014pva, Coloma:2016gei}.  Such
degeneracy makes it impossible to determine the neutrino mass ordering
by oscillation experiments alone~\cite{Coloma:2016gei} and therefore
jeopardizes one of the main goals of the upcoming neutrino oscillation
program.  Although the degeneracy is not exact when including
oscillation data from neutrino propagating in different environments,
quantitatively the breaking is small and global fits show that the
LMA-D solution is still pretty much allowed by oscillation data
alone~\cite{Gonzalez-Garcia:2013usa, Esteban:2018ppq}.  A second
important limitation of oscillation data is that it is only sensitive
to differences in the potential felt by different neutrino flavours.
Thus, oscillation data may be used to derive constraints on the
differences between diagonal NC NSI parameters, but not on the
individual parameters themselves.

Neutrino scattering data is also sensitive to NC NSI as they would
affect the interaction rates directly~\cite{Davidson:2003ha,
  Scholberg:2005qs, Barranco:2005ps, Barranco:2005yy, Barranco:2007ej,
  Bolanos:2008km}.  Therefore the combination of oscillation and
scattering data may be used to break the LMA-D degeneracy (see for
example Refs.~\cite{Miranda:2004nb, Escrihuela:2009up} for early works
on this topic).  However, such a combination is most effective
whenever the scattering cross section also falls in the
contact-interaction regime, since otherwise bounds obtained from
scattering may be evaded for sufficiently light
mediators~\cite{Farzan:2015doa}.  As a consequence, for NSI with
quarks the most powerful constraints come from measurements of
Coherent Elastic Neutrino-Nucleus Scattering~\cite{Freedman:1973yd}
(CE$\nu$NS), since the momentum transfer is small and therefore apply
to a wider class of models, as discussed in
Refs.~\cite{Coloma:2016gei, Coloma:2017egw, Shoemaker:2017lzs}.
Reference~\cite{Coloma:2017ncl} showed the impact obtained from the
joint analysis of CE$\nu$NS and oscillation data, considering NSI with
only one quark type at a time and using the first measurement of
CE$\nu$NS at COHERENT~\cite{COHERENT:2017ipa}.  Our later
works~\cite{Esteban:2018ppq, Coloma:2019mbs} subsequently improved
over this first analysis by including the latest oscillation data,
refining the treatment of COHERENT data~\cite{COHERENT:2018imc} and,
most importantly, by allowing for NC NSI with up and down quarks
simultaneously in the fit, which made the results more general.
Moreover, besides lifting the degeneracy, the addition of oscillation
and scattering data also allows to obtain separate constraints on the
diagonal NSI parameters~\cite{Coloma:2017ncl}.

The field of CE$\nu$NS is moving fast.  Since the first measurements
on CsI, the COHERENT collaboration has reported a separate measurement
using an Ar detector~\cite{COHERENT:2020iec} and the Dresden-II
experiment recently reported a signal using a Ge
detector~\cite{Colaresi:2021kus}.  The combination of CE$\nu$NS
measurements for different target nuclei is highly relevant to lift
the LMA-D degeneracy for NC NSI with arbitrary couplings to up and
down quarks~\cite{Scholberg:2005qs, Barranco:2005yy, Coloma:2017egw,
  Chaves:2021pey}.  Additionally, the combination of data obtained
using neutrinos from spallation sources and from reactors may be used
to lift degeneracies in NSI flavour space~\cite{Coloma:2022avw}.
Therefore, a significant improvement on the bounds for NC NSI with
quarks may be expected from the addition of the new datasets that have
become recently available.

As mentioned above, the global analyses in
Refs.~\cite{Gonzalez-Garcia:2013usa, Coloma:2017ncl, Esteban:2018ppq,
  Coloma:2019mbs} only included NSI with quarks.  However, there is no
reason to avoid the new current to couple to electrons as well.  The
presence of NSI with electrons would affect not only neutrino
propagation but also the interaction cross-section for electron
scattering (ES).  This would impact the data at both SK and Borexino,
and makes the problem much more demanding from the computational point
of view.  Constraints on NC NSI with electrons were obtained from the
analysis of Borexino Phase-II spectrum by the Borexino
Collaboration~\cite{Borexino:2019mhy} assuming only one NC NSI
coupling at a time.  Recently, in Ref.~\cite{Coloma:2022umy} we
performed an analysis of the Borexino Phase-II spectral data including
all NC NSI operators involving electrons simultaneously in the fit.
Our results showed that the simultaneous presence of several operators
significantly deteriorates the resulting bounds.  However, the
sensitivity of Borexino to NC NSI with electrons stems mainly from
their impact on the detection cross section and not from oscillations.
Therefore a significant improvement on the final constraints would be
expected from the combination with oscillation data from other
experiments.  While partial analyses for electron NSI using a subset
of data have been performed before~\cite{Miranda:2004nb,
  Escrihuela:2009up, Bolanos:2008km}, an analysis using all available
neutrino oscillation data has not been performed yet in this context.
Furthermore, the presence of the LMA-D solution in presence of NSI
with electrons remains an open question.

In this work we address these open issues by updating and extending
our analyses in Refs.~\cite{Esteban:2018ppq, Coloma:2019mbs}
accounting for the possibility of NC NSI with up quarks, down quarks
and electrons \emph{simultaneously}.  We will consider vector and
axial-vector couplings separately.  Our global analysis updates that
of our previous works, with the major addition of the Borexino
Phase-II spectral dataset.  Furthermore when combining with CE$\nu$NS
we include the results from COHERENT data on
CsI~\cite{COHERENT:2017ipa, COHERENT:2018imc} and Ar
targets~\cite{COHERENT:2020iec, COHERENT:2020ybo} together with the
recent results from CE$\nu$NS searches using reactor neutrinos at
Dresden-II reactor experiment~\cite{Colaresi:2022obx, Coloma:2022avw}.
To this aim, in Sec.~\ref{sec:formalism} we briefly summarize the
framework of our study.  We present the main ingredients and
assumptions in the analysis of the NSI effects in the matter potential
in atmospheric and long-baseline (LBL) experiments in
Sec.~\ref{sec:formalism-earth}, and Solar and KamLAND in
Sec.~\ref{sec:formalism-solar}.  In particular, we include a
discussion on the possible departures from adiabaticity of the
evolution of solar neutrinos in Sec.~\ref{sec:formADIA}.  The effects
of NC NSI on the detection cross sections and its interplay with
flavour transitions in propagation is reviewed in
Sec.~\ref{sec:formCS} including the specific modification of the cross
section for ES (Sec.~\ref{sec:formES}), NC scattering with quarks
(Sec.~\ref{sec:sno-nc}), and CE$\nu$NS (Sec.~\ref{sec:formCNUES}).
The results of the different analyses performed are presented in
Sec.~\ref{sec:results} where we derive our current knowledge of the
size and flavour structure of vector NC NSI coupling to either
electrons (Sec.~\ref{sec:resule}), up and down quarks
(Sec.~\ref{sec:resulq}), or a general combination of those
(Sec.~\ref{sec:resulgen}).  We study the status of the LMA-D solution
as function of the relative strength of the non-standard couplings to
the different charged fermions in Sec.~\ref{sec:resulLMAD}.  We also
quantify the results as allowed ranges of the effective combinations
of couplings relevant for long-baseline and atmospheric oscillation
experiments.  In Sec.~\ref{sec:summary} we summarize our conclusions.

\section{Formalism}
\label{sec:formalism}

As stated in the introduction, in this work we will consider NC NSI
(hereafter referred to simply as NSI, for brevity) relevant to
neutrino scattering and propagation in matter, as parametrized in
Eq.~\eqref{eq:nsi-nc}.  To make the analysis feasible, the following
simplifications are introduced:
\begin{itemize}
\item we assume that the neutrino flavour structure of the interactions
  is independent of the charged fermion properties;

\item we further assume that the chiral structure of the charged
  fermion vertex is the same for all fermion types.
\end{itemize}
Under these hypotheses, we can factorize $\Eps_{\alpha\beta}^{f,P}$ as
the product of three terms:
\begin{equation}
  \label{eq:eps-fact}
  \Eps_{\alpha\beta}^{f,P}
  \equiv \Eps_{\alpha\beta} \, \xi^f \chi^P
\end{equation}
where the matrix $\Eps_{\alpha\beta}$ describes the dependence on the
neutrino flavour, the coefficients $\xi^f$ parametrize the coupling to
the charged fermions, and the terms $\chi^P$ account for the chiral
structure of such couplings, normalized so that
$\Eps_{\alpha\beta}^{f,L}$ ($\Eps_{\alpha\beta}^{f,R}$) corresponds to
$\chi^L = 1/2$ and $\chi^R = 0$ ($\chi^R = 1/2$ and $\chi^L = 0$).
With these assumptions the Lagrangian in Eq.~\eqref{eq:nsi-nc} takes
the form:
\begin{equation}
  \mathcal{L}_\text{NSI,NC} = -2\sqrt{2} G_F
  \bigg[ \sum_{\alpha,\beta} \Eps_{\alpha\beta}
  (\bar\nu_\alpha\gamma^\mu P_L\nu_\beta) \bigg]
  \bigg[ \sum_{f,P} \xi^f \chi^P (\bar f\gamma_\mu P f) \bigg] \,.
\end{equation}
Concerning the chiral structure of the charged fermion vertex, in this
work we will consider either vector couplings ($\chi^L = \chi^R =
1/2$), or axial-vector couplings ($\chi^L = -\chi^R = 1/2)$.  The
corresponding combinations of NSI coefficients are given in
Eq.~\eqref{eq:vect-axial}.

For what concerns the dependence of the NSI on the charged fermion
type, we notice that ordinary matter is composed of electrons ($e$),
up quarks ($u$) and down quarks ($d$), so that only the coefficients
$\xi^e$, $\xi^u$ and $\xi^d$ are experimentally accessible.  For
vector NSI, since quarks are always confined inside protons ($p$) and
neutrons ($n$), we may define (see, \textit{e.g.},
Ref.~\cite{Breso-Pla:2023tnz}):
\begin{equation}
  \label{eq:xi-nucleon}
  \xi^p = 2\xi^u + \xi^d \,,
  \qquad
  \xi^n = 2\xi^d + \xi^u
\end{equation}
so that $\Eps_{\alpha\beta}^{p,V} \equiv 2\Eps_{\alpha\beta}^{u,V} +
\Eps_{\alpha\beta}^{d,V} = \Eps_{\alpha\beta}\, \xi^p\, (\chi^L +
\chi^R)$ and $\Eps_{\alpha\beta}^{n,V} \equiv
2\Eps_{\alpha\beta}^{d,V} + \Eps_{\alpha\beta}^{u,V} =
\Eps_{\alpha\beta}\, \xi^n\, (\chi^L + \chi^R)$.
For axial-vector NSI, the correspondence between quark NSI and nucleon
NSI is not that obvious: for example, for non-relativistic nucleons an
axial-vector hadronic current would induce a change in the spin of the
nucleon.  In our analysis this type of NSI is only relevant for the
breakup of deuterium at SNO as discussed in Sec.~\ref{sec:sno-nc}.
In any case, it is clear that a simultaneous re-scaling of all
$\{\xi^f\}$ by a common factor can be reabsorbed into a re-scaling of
$\Eps_{\alpha\beta}$, so that only the direction of $(\xi^e, \xi^u,
\xi^d)$ --~or, equivalently, $(\xi^e, \xi^p, \xi^n)$~-- is
phenomenologically non-trivial.  We parametrize such direction using
spherical coordinates, in terms of a ``latitude'' angle $\eta$ and a
``longitude'' angle $\zeta$, as an extension of the framework and
notation introduced in Ref.~\cite{Esteban:2018ppq} (see also
Ref.~\cite{Amaral:2023tbs}).  Concretely, we define:
\begin{equation}
  \label{eq:xi-eta}
  \xi^e = \sqrt{5} \cos\eta \sin\zeta \,,
  \qquad
  \xi^p = \sqrt{5} \cos\eta \cos\zeta \,,
  \qquad
  \xi^n = \sqrt{5} \sin\eta
\end{equation}
or, in terms of the ``quark'' couplings:
\begin{equation}
  \label{eq:xi-eta-quark}
  \xi^u = \frac{\sqrt{5}}{3} (2 \cos\eta \cos\zeta - \sin\eta) \,,
  \qquad
  \xi^d = \frac{\sqrt{5}}{3} (2 \sin\eta - \cos\eta \cos\zeta)
\end{equation}
Using this parametrization, the case of NSI with quarks analyzed in
Ref.~\cite{Esteban:2018ppq} correspond to the ``prime meridian''
$\zeta = 0$, with the pure up-quark and pure down-quark cases located
at $\eta = \arctan(1/2) \approx 26.6^\circ$ and $\eta = \arctan(2)
\approx 63.4^\circ$, respectively.  The ``poles'' ($\eta = \pm
90^\circ$) correspond to NSI with neutrons, while NSI with protons
lies on the ``equator'' ($\eta = \zeta = 0$).  Finally, the pure
electron case is also on the equator, at a right angle ($\zeta = \pm
90^\circ$) from the pure proton case.  Notice that an overall sign
flip of $(\xi^e, \xi^u, \xi^d)$ is just a special case of re-scaling
and produces no observable effect, hence it is sufficient to restrict
both $\eta$ and $\zeta$ to the $[-90^\circ, +90^\circ]$ range.

The presence of vector NC NSI will affect both neutrino propagation in
matter and neutrino scattering in the detector, while axial-vector NC
NSI only affect some of the interaction cross sections.  Both
propagation and interaction effects lead to a modification of the
expected number of events which can be described by the generic
expression~\cite{Coloma:2022umy}:
\begin{equation}
  \label{eq:ES-dens}
  N_\text{ev} \propto
  \Tr\Big[ \rho^\text{det}\, \sigma^\text{det} \Big]
\end{equation}
where $\rho^\text{det}$ is the density matrix characterizing the
flavour state of the neutrinos reaching the detector, while the
\emph{generalized} cross section $\sigma^\text{det}$ is a matrix in
flavour space containing enough information to describe the
interaction of \emph{any} neutrino configuration.  The form of
Eq.~\eqref{eq:ES-dens} is manifestly basis-independent and permits a
separate description of propagation effects (encoded into
$\rho^\text{det}$) and of the scattering process (contained in
$\sigma^\text{det}$), while at the same time properly taking into
account possible interference between them.  Notice that both
$\rho^\text{det}$ and $\sigma^\text{det}$ are hermitian matrices,
which ensures that $N_\text{ev}$ is real.  Actually,
Eq.~\eqref{eq:ES-dens} is invariant under the joint transformation:
\begin{equation}
  \label{eq:conjugate}
  \rho^\text{det} \to \big[ \rho^\text{det} \big]^*
  \quad\text{and}\quad
  \sigma^\text{det} \to \big[ \sigma^\text{det} \big]^*
\end{equation}
whose implications will be discussed later on in this section.

\subsection{Neutrino oscillations in the presence of NSI}
\label{sec:formOSC}

In general, the evolution of the neutrino and antineutrino flavour
state during propagation is governed by the Hamiltonian:
\begin{equation}
  H^\nu = H_\text{vac} + H_\text{mat}
  \quad\text{and}\quad
  H^{\bar\nu} = ( H_\text{vac} - H_\text{mat} )^* \,,
\end{equation}
where $H_\text{vac}$ is the vacuum part which in the flavour basis
$(\nu_e, \nu_\mu, \nu_\tau)$ reads
\begin{equation}
  \label{eq:Hvac}
  H_\text{vac} = U_\text{vac} D_\text{vac} U_\text{vac}^\dagger
  \quad\text{with}\quad
  D_\text{vac} = \frac{1}{2E_\nu} \diag(0, \Dmq_{21}, \Dmq_{31}) \,.
\end{equation}
Here $U_\text{vac}$ denotes the three-lepton mixing matrix in
vacuum~\cite{Pontecorvo:1967fh, Maki:1962mu, Kobayashi:1973fv}.
Following the convention of Ref.~\cite{Coloma:2016gei}, we define
$U_\text{vac} = R_{23}(\theta_{23}) R_{13}(\theta_{13})
\tilde{R}_{12}(\theta_{12}, \delta_\text{CP})$, where
$R_{ij}(\theta_{ij})$ is a rotation of angle $\theta_{ij}$ in the $ij$
plane and $\tilde{R}_{12}(\theta_{12},\delta_\text{CP})$ is a complex
rotation by angle $\theta_{12}$ and phase $\delta_\text{CP}$.
Explicitly:
\begin{equation}
  \label{eq:Uvac}
  U_\text{vac} =
  \begin{pmatrix}
    c_{12} c_{13}
    & s_{12} c_{13} e^{i\delta_\text{CP}}
    & s_{13}
    \\
    - s_{12} c_{23} e^{-i\delta_\text{CP}} - c_{12} s_{13} s_{23}
    & \hphantom{+} c_{12} c_{23} - s_{12} s_{13} s_{23} e^{i\delta_\text{CP}}
    & c_{13} s_{23}
    \\
    \hphantom{+} s_{12} s_{23} e^{-i\delta_\text{CP}} - c_{12} s_{13} c_{23}
    & - c_{12} s_{23} - s_{12} s_{13} c_{23} e^{i\delta_\text{CP}}
    & c_{13} c_{23}
  \end{pmatrix}
\end{equation}
where $c_{ij} \equiv \cos\theta_{ij}$ and $s_{ij} \equiv
\sin\theta_{ij}$.  This expression differs from the usual one ``$U$''
(defined, \textit{e.g.}, in Eq.~(1.1) of Ref.~\cite{Esteban:2016qun})
by an overall phase matrix: $U_\text{vac} = P U P^*$ with $P =
\diag(e^{i\delta_\text{CP}}, 1, 1)$.  It is easy to show that, in the
absence of NSI, such rephasing produces no visible effect, so that when
only standard interactions are considered the physical interpretation
of the vacuum parameters ($\Dmq_{21}$, $\Dmq_{31}$, $\theta_{12}$,
$\theta_{13}$, $\theta_{23}$ and $\delta_\text{CP}$) is exactly the
same in both conventions.
The advantage of defining $U_\text{vac}$ as in Eq.~\eqref{eq:Uvac} is
that the transformation $H_\text{vac} \to -H_\text{vac}^*$, whose
relevance for the present work will be discussed below, can be
implemented exactly (up to an irrelevant multiple of the identity) by
the following transformation of the parameters:
\begin{equation}
  \label{eq:osc-deg}
  \begin{aligned}
    \Dmq_{31} &\to -\Dmq_{31} + \Dmq_{21} = -\Dmq_{32} \,,
    \\
    \theta_{12} & \to \pi/2 - \theta_{12} \,,
    \\
    \delta_\text{CP} &\to \pi - \delta_\text{CP}
  \end{aligned}
\end{equation}
which does not spoil the commonly assumed restrictions on the range of
the vacuum parameters ($\Dmq_{21} > 0$ and $0 \leq \theta_{ij} \leq
\pi/2$).

Concerning the matter part $H_\text{mat}$ of the Hamiltonian which
governs neutrino oscillations, if all possible operators in
Eq.~\eqref{eq:nsi-nc} are added to the SM Lagrangian we get:
\begin{equation}
  \label{eq:Hmat}
  H_\text{mat} = \sqrt{2} G_F N_e(x)
  \begin{pmatrix}
    1+\Epx_{ee}(x) & \Epx_{e\mu}(x) & \Epx_{e\tau}(x) \\
    \Epx_{e\mu}^*(x) & \Epx_{\mu\mu}(x) & \Epx_{\mu\tau}(x) \\
    \Epx_{e\tau}^*(x) & \Epx_{\mu\tau}^*(x) & \Epx_{\tau\tau}(x)
  \end{pmatrix}
\end{equation}
where the ``$+1$'' term in the $ee$ entry accounts for the standard
contribution, and
\begin{equation}
  \label{eq:epx-nsi}
  \Epx_{\alpha\beta}(x) = \sum_{f=e,u,d}
  \frac{N_f(x)}{N_e(x)} \Eps_{\alpha\beta}^{f,V}
\end{equation}
describes the non-standard part.  Here $N_f(x)$ is the number density
of fermion $f$ as a function of the distance traveled by the neutrino
along its trajectory.  In Eq.~\eqref{eq:epx-nsi} we have limited the
sum to the charged fermions present in ordinary matter, $f=e,u,d$.
Taking into account that $N_u(x) = 2N_p(x) + N_n(x)$ and $N_d(x) =
N_p(x) + 2N_n(x)$, and also that matter neutrality implies $N_p(x) =
N_e(x)$, Eq.~\eqref{eq:epx-nsi} becomes:
\begin{equation}
  \label{eq:epx-nuc}
  \Epx_{\alpha\beta}(x) =
  \big( \Eps_{\alpha\beta}^{e,V} + \Eps_{\alpha\beta}^{p,V} \big)
  + Y_n(x)\, \Eps_{\alpha\beta}^{n,V}
  \quad\text{with}\quad
  Y_n(x) \equiv \frac{N_n(x)}{N_e(x)}
\end{equation}
which shows that, from the phenomenological point of view, the
propagation effects of NSI with electrons can be mimicked by NSI with
quarks by means of a suitable combination of up-quark and down-quark
contributions.

Since the matter potential can be determined by oscillation
experiments only up to an overall multiple of the identity, each
$\Eps_{\alpha\beta}^{f,V}$ matrix introduces 8 new parameters: two
differences of the three diagonal real parameters (\textit{e.g.},
$\Eps_{ee}^{f,V} - \Eps_{\mu\mu}^{f,V}$ and $\Eps_{\tau\tau}^{f,V} -
\Eps_{\mu\mu}^{f,V}$) and three off-diagonal complex parameters
(\textit{i.e.}, three additional moduli and three complex phases).
Under the assumption that the neutrino flavour structure of the
interactions is independent of the charged fermion type, as described
in Eq.~\eqref{eq:eps-fact}, we get:
\begin{equation}
  \label{eq:epx-eta}
  \begin{aligned}
    \Epx_{\alpha\beta}(x)
    &= \Eps_{\alpha\beta} \big[ \xi^e + \xi^p + Y_n(x) \xi^n \big]
    \big( \chi^L + \chi^R \big)
    \\
    &= \sqrt{5} \, \big[\! \cos\eta\, (\cos\zeta + \sin\zeta)
      + Y_n(x) \sin\eta \big] \big( \chi^L + \chi^R \big)\,
    \Eps_{\alpha\beta}
  \end{aligned}
\end{equation}
so that the phenomenological framework adopted here is characterized
by 10 matter parameters: eight related to the matrix
$\Eps_{\alpha\beta}$ plus two directions $(\eta, \zeta)$ in the
$(\xi^e, \xi^p, \xi^n)$ space.  Notice, however, that the dependence
on $\zeta$ in Eq.~\eqref{eq:epx-eta} can be reabsorbed into a
re-scaling of $\Eps_{\alpha\beta}$ by introducing a new effective angle
$\eta^\prime$:
\begin{multline}
  \label{eq:epx-etapr}
  \Epx_{\alpha\beta}(x)
  = \sqrt{5} \, \big[\! \cos\eta^\prime
    + Y_n(x) \sin\eta^\prime \big] \big( \chi^L + \chi^R \big)\,
  \Eps_{\alpha\beta}^\prime
  \\
  \text{with}\quad
  \tan\eta^\prime
  \equiv \tan\eta \mathbin{\big/} (\cos\zeta + \sin\zeta)
  \quad\text{and}\quad
  \Eps_{\alpha\beta}^\prime
  \equiv \Eps_{\alpha\beta} \sqrt{1 + \cos^2\eta \sin(2\zeta)} \,.
\end{multline}
This is a consequence of the fact that electron and proton NSI always
appear together in propagation, as explained after
Eq.~\eqref{eq:epx-nuc}.  Indeed, $\eta^\prime$ is just a practical way
to express the direction in the $(\xi^e + \xi^p,\, \xi^n)$ plane.  For
$\zeta = 0$, which is the case studied in Ref.~\cite{Esteban:2018ppq},
one trivially recovers $\eta^\prime = \eta$ and
$\Eps_{\alpha\beta}^\prime = \Eps_{\alpha\beta}$.
Furthermore, for $\zeta = -45^\circ$ one gets $\xi^e + \xi^p = 0$, so
that NSI effects in oscillations depend solely on the neutron coupling
$\xi^n$.
This would be the case, for example, in models where the $Z'$ does not
couple directly to matter fermions, and NSI are generated through
$Z-Z'$ mass mixing (see the related discussion in
Ref.~\cite{Heeck:2018nzc}).  In fact, in this case, the specific value
of $\eta$ becomes irrelevant as long as it is different from zero
(since Eq.~\eqref{eq:epx-etapr} always yields $\eta^\prime = \pm
90^\circ$ in this case), while for the special point where $\eta = 0$
and $\zeta = -45^\circ$ NSI completely cancel from oscillations.  It
should be stressed, however, that this only applies to
\emph{oscillations}: the implications of NSI for neutrino scattering
described in Sec.~\ref{sec:formCS} will still depend non-trivially on
$\eta$.

We finish this section by reminding that the neutrino transition
probabilities remain invariant --~and more generically the density
matrix $\rho^\text{det}$ undergoes complex conjugation, as described
in Eq.~\eqref{eq:conjugate}~-- if the Hamiltonian $H^\nu =
H_\text{vac} + H_\text{mat}$ is transformed as $H^\nu \to -(H^\nu)^*$.
This requires a simultaneous transformation of both the vacuum and the
matter terms.  The transformation of $H_\text{vac}$ is described in
Eq.~\eqref{eq:osc-deg} and involves a change in the octant of
$\theta_{12}$ (the so-called LMA-D~\cite{Miranda:2004nb} solution) as
well as a change in the neutrino mass ordering (\textit{i.e.}, the
sign of $\Dmq_{31}$), which is why it has been called ``generalized
mass-ordering degeneracy'' in Ref.~\cite{Coloma:2016gei}.  As for
$H_\text{mat}$ we need:
\begin{equation}
  \label{eq:NSI-deg}
  \begin{aligned}
    \big[ \Epx_{ee}(x) - \Epx_{\mu\mu}(x) \big]
    &\to - \big[ \Epx_{ee}(x) - \Epx_{\mu\mu}(x) \big] - 2 \,,
    \\
    \big[ \Epx_{\tau\tau}(x) - \Epx_{\mu\mu}(x) \big]
    &\to -\big[ \Epx_{\tau\tau}(x) - \Epx_{\mu\mu}(x) \big] \,,
    \\
    \Epx_{\alpha\beta}(x)
    &\to - \Epx_{\alpha\beta}^*(x) \qquad (\alpha \neq \beta) \,,
  \end{aligned}
\end{equation}
see Refs.~\cite{Gonzalez-Garcia:2013usa, Bakhti:2014pva,
  Coloma:2016gei}.  As seen in Eqs.~\eqref{eq:epx-nsi},
\eqref{eq:epx-nuc} and~\eqref{eq:epx-eta} the matrix
$\Epx_{\alpha\beta}(x)$ depends on the chemical composition of the
medium, which may vary along the neutrino trajectory, so that in
general the condition in Eq.~\eqref{eq:NSI-deg} is fulfilled only in
an approximate way.  The degeneracy becomes exact in the following two
cases:\footnote{Strictly speaking, Eq.~\eqref{eq:NSI-deg} can be
satisfied exactly for \emph{any} matter chemical profile $Y_n(x)$ if
$\Eps_{\alpha\beta}^{n,V}$ and $\Eps_{\alpha\beta}^{e,V} +
\Eps_{\alpha\beta}^{p,V}$ are allowed to transform independently of
each other.  This possibility, however, is incompatible with the
factorization constraint of Eq.~\eqref{eq:eps-fact}, so it will not be
discussed here.}
\begin{itemize}
\item if the effective NSI coupling to neutrons vanishes, so that
  $\Eps_{\alpha\beta}^{n,V} = 0$ in Eq.~\eqref{eq:epx-nuc}.  In terms
  of fundamental quantities this occurs when $\Eps_{\alpha\beta}^{u,V}
  = -2 \Eps_{\alpha\beta}^{d,V}$, \textit{i.e.}, the NSI couplings are
  proportional to the electric charge of quarks.  In our
  parametrization this corresponds to the ``equator'' $\eta=0$ for
  arbitrary $\zeta$, as shown in Eq.~\eqref{eq:epx-eta};

\item if the neutron/proton ratio $Y_n(x)$ is constant along the
  entire neutrino propagation path.  This is certainly the case for
  reactor and long-baseline experiments, where only the Earth's mantle
  is involved, and to a good approximation also for atmospheric
  neutrinos, since the differences in chemical composition between
  mantle and core can safely be neglected in the context of
  NSI~\cite{GonzalezGarcia:2011my}.  In this case the matrix
  $\Epx_{\alpha\beta}(x)$ becomes independent of $x$ and can be
  regarded as a new phenomenological parameter, as we will describe in
  Sec.~\ref{sec:formalism-earth}.
\end{itemize}
Further details on the implications of this degeneracy for different
classes of neutrino experiments (solar, atmospheric, \textit{etc.}) is
provided below in the corresponding section.

\subsubsection{Matter potential in atmospheric and long-baseline neutrinos}
\label{sec:formalism-earth}

As discussed in Ref.~\cite{GonzalezGarcia:2011my}, in the Earth the
neutron/proton ratio $Y_n(x)$ which characterizes the matter chemical
composition can be taken to be constant to very good approximation.
The PREM model~\cite{Dziewonski:1981xy} fixes $Y_n = 1.012$ in the
Mantle and $Y_n = 1.137$ in the Core, with an average value
$Y_n^\oplus = 1.051$ all over the Earth.  Setting therefore $Y_n(x)
\equiv Y_n^\oplus$ in Eqs.~\eqref{eq:epx-nsi} and~\eqref{eq:epx-nuc}
we get $\Epx_{\alpha\beta}(x) \equiv \Eps_{\alpha\beta}^\oplus$ with:
\begin{equation}
  \label{eq:eps-earth0}
  \Eps_{\alpha\beta}^\oplus
  = \Eps_{\alpha\beta}^{e,V} + \big( 2 + Y_n^\oplus \big) \Eps_{\alpha\beta}^{u,V}
  + \big( 1 + 2Y_n^\oplus \big) \Eps_{\alpha\beta}^{d,V}
  = \big( \Eps_{\alpha\beta}^{e,V} + \Eps_{\alpha\beta}^{p,V} \big)
  + Y_n^\oplus \Eps_{\alpha\beta}^{n,V} \,.
\end{equation}
If we impose quark-lepton factorization as in Eq.~\eqref{eq:epx-eta}
we get:
\begin{equation}
  \label{eq:eps-earth}
  \begin{aligned}
    \Eps_{\alpha\beta}^\oplus
    &= \sqrt{5} \, \big[\! \cos\eta\, (\cos\zeta + \sin\zeta)
      + Y_n^\oplus \sin\eta \big] \big( \chi^L + \chi^R \big)\,
    \Eps_{\alpha\beta}
    \\
    &= \sqrt{5} \, \big[\! \cos\eta^\prime
      + Y_n^\oplus \sin\eta^\prime \big]
    \big( \chi^L + \chi^R \big)\, \Eps_{\alpha\beta}^\prime \,.
  \end{aligned}
\end{equation}
In other words, within this approximation the analysis of atmospheric
and LBL neutrinos holds for any combination of NSI with up, down or
electrons and it can be performed in terms of the effective NSI
couplings $\Eps_{\alpha\beta}^\oplus$, which play the role of
phenomenological parameters.  In particular, the best-fit value and
allowed ranges of $\Eps_{\alpha\beta}^\oplus$ are independent of
$\eta$ and $\zeta$, while the bounds on $\Eps_{\alpha\beta}$ simply
scale as $[\cos\eta\, (\cos\zeta + \sin\zeta) + Y_n^\oplus \sin\eta]$.
Moreover, it is immediate to see that for $\eta^\prime =
\arctan(-1/Y_n^\oplus) \approx -43.6^\circ$, with $\eta^\prime$
defined in Eq.~\eqref{eq:epx-etapr}, the contribution of NSI to the
matter potential vanishes, so that no bound on $\Eps_{\alpha\beta}$
can be derived from atmospheric and LBL data in such case.
This would be approximately the case, for example, for $U(1)'$ models
associated to the combination $B - 2L_e + \alpha L_\mu - \beta L_\tau$
and, consequently, oscillation bounds are significantly relaxed for
this type of models~\cite{Greljo:2022dwn}.

Following the approach of Ref.~\cite{GonzalezGarcia:2011my}, the
matter Hamiltonian $H_\text{mat}$, given in Eq.~\eqref{eq:Hmat} after
setting $\Epx_{\alpha\beta}(x) \equiv \Eps_{\alpha\beta}^\oplus$, can
be parametrized in a way that mimics the structure of the vacuum
term~\eqref{eq:Hvac}:
\begin{equation}
  \label{eq:HmatGen}
  H_\text{mat} = Q_\text{rel} U_\text{mat} D_\text{mat}
  U_\text{mat}^\dagger Q_\text{rel}^\dagger
  \text{~~with~~}
  \left\lbrace
  \begin{aligned}
    Q_\text{rel} &= \diag\left(
    e^{i\alpha_1}, e^{i\alpha_2}, e^{-i\alpha_1 -i\alpha_2} \right),
    \\
    U_\text{mat} &= R_{12}(\varphi_{12}) R_{13}(\varphi_{13})
    \tilde{R}_{23}(\varphi_{23}, \delta_\text{NS}) \,,
    \\
    D_\text{mat} &= \sqrt{2} G_F N_e(x)
    \diag(\Eps_\oplus, \Eps_\oplus^\prime, 0)
  \end{aligned}\right.
\end{equation}
where $R_{ij}(\varphi_{ij})$ is a rotation of angle $\varphi_{ij}$ in
the $ij$ plane and $\tilde{R}_{23}(\varphi_{23},\delta_\text{NS})$ is
a complex rotation by angle $\varphi_{23}$ and phase
$\delta_\text{NS}$.
Note that the two phases $\alpha_1$ and $\alpha_2$ included in
$Q_\text{rel}$ are not a feature of neutrino-matter interactions, but
rather a relative feature of the vacuum and matter terms.
In order to simplify the analysis we impose that two eigenvalues of
$H_\text{mat}$ are equal, $\Eps_\oplus^\prime = 0$.  This assumption
is justified since, as shown in Ref.~\cite{Friedland:2004ah}, in this
case strong cancellations in the oscillation of atmospheric neutrinos
occur, and this is precisely the situation in which the weakest
constraints can be placed.
Setting $\Eps_\oplus^\prime \to 0$ implies that the $\varphi_{23}$
angle and the $\delta_\text{NS}$ phase disappear from neutrino
oscillations, so that the effective NSI couplings
$\Eps_{\alpha\beta}^\oplus$ can be parametrized as:
\begin{equation}
  \label{eq:eps_atm}
  \begin{aligned}
    \Eps_{ee}^\oplus - \Eps_{\mu\mu}^\oplus
    &= \hphantom{-} \Eps_\oplus \, (\cos^2\varphi_{12} - \sin^2\varphi_{12})
    \cos^2\varphi_{13} - 1\,,
    \\
    \Eps_{\tau\tau}^\oplus - \Eps_{\mu\mu}^\oplus
    &= \hphantom{-} \Eps_\oplus \, (\sin^2\varphi_{13}
    - \sin^2\varphi_{12} \, \cos^2\varphi_{13}) \,,
    \\
    \Eps_{e\mu}^\oplus
    &= -\Eps_\oplus \, \cos\varphi_{12} \, \sin\varphi_{12} \,
    \cos^2\varphi_{13} \, e^{i(\alpha_1 - \alpha_2)} \,,
    \\
    \Eps_{e\tau}^\oplus
    &= -\Eps_\oplus \, \cos\varphi_{12} \, \cos\varphi_{13} \,
    \sin\varphi_{13} \, e^{i(2\alpha_1 + \alpha_2)} \,,
    \\
    \Eps_{\mu\tau}^\oplus
    &= \hphantom{-} \Eps_\oplus \, \sin\varphi_{12} \, \cos\varphi_{13} \,
    \sin\varphi_{13} \, e^{i(\alpha_1 + 2\alpha_2)} \,.
  \end{aligned}
\end{equation}
As further simplification, in order to keep the fit manageable we
assume real NSI, which we implement by choosing $\alpha_1 = \alpha_2 =
0$ and $-\pi/2 \leq \varphi_{ij} \leq \pi/2$, and also restrict
$\delta_\text{CP} \in \{0, \pi\}$.  It is important to note that with
these approximations the formalism for atmospheric and long-baseline
data is CP-conserving; we will go back to this point when discussing
the experimental results included in the fit.  In addition to
atmospheric and LBL experiments, important information on neutrino
oscillation parameters is provided also by reactor experiments with a
baseline of about 1~km.  Due to the very small amount of matter
crossed, both standard and non-standard matter effects are completely
irrelevant for these experiments, so that neutrino propagation depends
only on the vacuum parameters.

\subsubsection{Matter potential for solar and KamLAND neutrinos}
\label{sec:formalism-solar}

For the study of propagation of solar and KamLAND neutrinos one can
work in the one mass dominance approximation, $\Dmq_{31} \to \infty$
(which effectively means that $G_F \sum_f N_f(x)\,
\Eps_{\alpha\beta}^{f,V} \ll \Dmq_{31} / E_\nu$).  In this limit the
neutrino evolution can be calculated in an effective $2\times 2$ model
described by the Hamiltonian $H_\text{eff} = H_\text{vac}^\text{eff} +
H_\text{mat}^\text{eff}$, with:
\begin{align}
  \label{eq:HvacSol}
  H_\text{vac}^\text{eff}
  &= \frac{\Dmq_{21}}{4 E_\nu}
  \begin{pmatrix}
    -\cos2\theta_{12} \, \hphantom{e^{-i\delta_\text{CP}}}
    & ~\sin2\theta_{12} \, e^{i\delta_\text{CP}}
    \\
    \hphantom{-}\sin2\theta_{12} \, e^{-i\delta_\text{CP}}
    & ~\cos2\theta_{12} \, \hphantom{e^{i\delta_\text{CP}}}
  \end{pmatrix},
  \\
  \label{eq:HmatSol}
  H_\text{mat}^\text{eff}
  &= \sqrt{2} G_F N_e(x)
  \left[
    \frac{c_{13}^2}{2}
    \begin{pmatrix}
      1 & \hphantom{-}0 \\
      0 & -1
    \end{pmatrix}
    + \big[ \xi^e + \xi^p + Y_n(x) \xi^n \big]
    \big( \chi^L + \chi^R \big)\!
    \begin{pmatrix}
      -\Eps_D^{\hphantom{*}} & \Eps_N \\
      \hphantom{+} \Eps_N^* & \Eps_D
    \end{pmatrix}
    \right],
\end{align}
where we have imposed the quark-lepton factorization of
Eq.~\eqref{eq:epx-eta} and used the parametrization convention of
Eq.~\eqref{eq:Uvac} for $U_\text{vac}$.  The coefficients $\Eps_D$ and
$\Eps_N$ are related to the original parameters $\Eps_{\alpha\beta}$
by the following relations:
\begin{align}
  \label{eq:eps_D}
  \begin{split}
    \Eps_D
    &= c_{13} s_{13}\, \Re\!\big( s_{23} \, \Eps_{e\mu}
    + c_{23} \, \Eps_{e\tau} \big)
    - \big( 1 + s_{13}^2 \big)\, c_{23} s_{23}\,
    \Re\!\big( \Eps_{\mu\tau} \big)
    \\
    & \hphantom{={}}
    -\frac{c_{13}^2}{2} \big( \Eps_{ee} - \Eps_{\mu\mu} \big)
    + \frac{s_{23}^2 - s_{13}^2 c_{23}^2}{2}
    \big( \Eps_{\tau\tau} - \Eps_{\mu\mu} \big) \,,
  \end{split}
  \\[2mm]
  \label{eq:eps_N}
  \Eps_N &=
  c_{13} \big( c_{23} \, \Eps_{e\mu} - s_{23} \, \Eps_{e\tau} \big)
  + s_{13} \left[
    s_{23}^2 \, \Eps_{\mu\tau} - c_{23}^2 \, \Eps_{\mu\tau}^*
    + c_{23} s_{23} \big( \Eps_{\tau\tau} - \Eps_{\mu\mu} \big)
    \right].
\end{align}
Denoting by $S_\text{eff}$ the $2\times 2$ unitary matrix obtained
integrating $H_\text{eff}$ along the neutrino trajectory, the full
density matrix $\rho^\text{det}$ introduced in Eq.~\eqref{eq:ES-dens}
can be written as:
\begin{equation}
  \rho_{\alpha\beta}^\text{det} = c_{13}^2 \big[
    A_{\alpha\beta} P_\text{osc}
    + B_{\alpha\beta} P_\text{int}
    + i C_{\alpha\beta} P_\text{ext} \big]
  + D_{\alpha\beta}
\end{equation}
where the effective probabilities $P_\text{osc}$, $P_\text{int}$ and
$P_\text{ext}$ are given by
\begin{equation}
  P_\text{osc} \equiv |S_{21}^\text{eff}|^2 \,,
  \qquad
  P_\text{int} \equiv \Re\big(
  S_{11}^\text{eff} S_{21}^{\text{eff}\,*} \big) \,,
  \qquad
  P_\text{ext} \equiv \Im\big(
  S_{11}^\text{eff} S_{21}^{\text{eff}\,*} \big) \,,
\end{equation}
and the numerical coefficients $A_{\alpha\beta}$, $B_{\alpha\beta}$,
$C_{\alpha\beta}$ and $D_{\alpha\beta}$ are defined as
\begin{equation}
  \begin{aligned}
    A_{\alpha\beta} &\equiv
    O_{\alpha 2} O_{\beta 2} - O_{\alpha 1} O_{\beta 1} \,,
    &\qquad
    B_{\alpha\beta} &\equiv
    O_{\alpha 1} O_{\beta 2} + O_{\alpha 2} O_{\beta 1} \,,
    \\
    C_{\alpha\beta} &\equiv
    O_{\alpha 1} O_{\beta 2} - O_{\alpha 2} O_{\beta 1} \,,
    &\qquad
    D_{\alpha\beta} &\equiv
    \sum_{i=\text{all}} O_{\alpha i} O_{\beta i} |O_{ei}|^2 \,.
  \end{aligned}
\end{equation}
with $O = R_{23}(\theta_{23}) R_{13}(\theta_{13})$.  Unlike in
Ref.~\cite{Esteban:2018ppq} where NSI did \emph{not} affect the
scattering process and only the $\rho_{ee}^\text{det}$ entry (which
depends exclusively on $P_\text{osc}$) was required, a rephasing of
$S_\text{eff}$ now produces visible consequences as it affects
$P_\text{int}$ and $P_\text{ext}$.  Moreover, $\nu_\mu$ and $\nu_\tau$
are no longer indistinguishable as their scattering amplitude may be
different under NSI, so that the $\theta_{23}$ angle acquires
relevance.  Hence, for each fixed value of $\eta$ and $\zeta$ the
density matrix $\rho^\text{det}$ for solar and KamLAND neutrinos
depends effectively on eight quantities: the four real oscillation
parameters $\theta_{12}$, $\theta_{13}$, $\theta_{23}$ and
$\Dmq_{21}$, the real $\Eps_D$ and complex $\Eps_N$ matter parameters,
and the CP phase $\delta_\text{CP}$.
As stated in Sec.~\ref{sec:formalism-earth} in this work we will
assume real NSI, implemented here by setting $\delta_\text{CP} \in
\{0, \pi\}$ and considering only real values for $\Eps_N$.

Unlike in the Earth, the matter chemical composition of the Sun varies
substantially along the neutrino trajectory, and consequently the
potential depends non-trivially on the specific combinations of
$(\xi^e + \xi^p)$ and $\xi^n$ couplings --~\textit{i.e.}, on the value
of $\eta^\prime$ as determined by the $(\eta, \zeta)$ parameters.
This implies that the generalized mass-ordering degeneracy is not
exact, except for $\eta = 0$ (in which case the NSI potential is
proportional to the standard MSW potential and an exact inversion of
the matter sign is possible).  However the transformation described in
Eqs.~\eqref{eq:osc-deg} and~\eqref{eq:NSI-deg} still results in a good
fit to the global analysis of oscillation data for a wide range of
values of $\eta^\prime$, and non-oscillation data are needed to break
this degeneracy~\cite{Coloma:2016gei}.

\subsubsection{Departures from adiabaticity in presence of NSI}
\label{sec:formADIA}

When computing neutrino evolution in the Sun, it is often assumed that
it takes place in the adiabatic regime, as this considerably
simplifies the calculation.  While this is the case for neutrino
oscillations within the LMA solution with a standard matter potential,
it is worthwhile asking if the inclusion of NP effects (such as NSI)
can lead to non-adiabatic transitions.  If this were the case, it
would require special care and may lead to interesting new
phenomenological consequences.  This possibility has been largely
overlooked in the literature, where most studies of NSI in the Sun
assume adiabatic transitions.

Let us consider the two-neutrino case, with a matter potential that
depends on the position $x$ along the neutrino path inside the Sun.
In the instantaneous mass basis, the Hamiltonian can be written as:
\begin{equation}
  \label{eq:instant-mass}
  i \frac{\dd}{\dd x}
  \begin{pmatrix}
    \tilde \nu_1 \\ \tilde \nu_2
  \end{pmatrix}
  = \begin{pmatrix}
    -\Delta_m(x) \enspace & -i \theta_m^\prime(x)
    \\
    i \theta_m^\prime(x) & \Delta_m(x)
  \end{pmatrix}
  \begin{pmatrix}
    \tilde \nu_1 \\ \tilde \nu_2
  \end{pmatrix}
\end{equation}
where $\theta_m(x)$ and $\Delta_m(x)$ refer to the mixing angle and
oscillation frequency in the presence of matter effects, and
$\theta^\prime(x) \equiv \dd\theta(x) \big/ \dd x$.  In this basis,
transitions between $\tilde \nu_1 \leftrightarrow \tilde \nu_2$ are
negligible as long as the off-diagonal terms $\theta_m^\prime(x)$ are
small compared to the diagonal entries $\Delta_m(x)$ in
Eq.~\eqref{eq:instant-mass}.  This leads to the so-called adiabaticity
condition, that is:
\begin{equation}
  \label{eq:adiabatic}
  \gamma^{-1}(x) \equiv
  \bigg| \frac{\theta_m^\prime(x)}{\Delta_m(x)} \bigg| \ll 1 \,.
\end{equation}
Neutrino transitions will be adiabatic if this condition is satisfied
along all points in the neutrino trajectory.  Of course, this argument
is general and may be applied both in the standard case and in the
presence of NP.  In the present work neutrino propagation is described
by the effective $2\times 2$ Hamiltionian $H_\text{eff}$ introduced in
Eqs.~\eqref{eq:HvacSol} and~\eqref{eq:HmatSol}.  Assuming
$\delta_\text{CP} = 0$ and real NSI, so that $H_\text{eff}$ is
traceless and real with $H_{11}^\text{eff} = -H_{22}^\text{eff}$ and
$H_{12}^\text{eff} = H_{21}^\text{eff}$, we can write:
\begin{equation}
  \theta_m(x)
  \equiv \frac{1}{2} \arctan\!\big[
    H_{12}^\text{eff}(x) \mathbin{\big/} H_{22}^\text{eff}(x) \big]
  \quad\text{and}\quad
  \Delta_m(x)
  \equiv \sqrt{\big[ H_{12}^\text{eff}(x) \big]^2
    + \big[ H_{22}^\text{eff}(x) \big]^2} \,.
\end{equation}
In the presence of NSI, for a given matter density and neutrino
energy, it is possible to choose $\Eps_D$ and $\Eps_N$ so that the
contribution from NP cancels the standard one, resulting in
$\Delta_m(x) \to 0$.  It is easy to show analytically that, for such
values, the adiabatic condition in Eq.~\eqref{eq:adiabatic} is no
longer satisfied.  Specifically, such cancellation takes place when:
\begin{equation}
  \label{eq:cancel}
  \begin{aligned}
    \big[ \xi^e + \xi^p + Y_n(x) \xi^n \big]
    \big(\chi^L + \chi^R \big)\, \Eps_D
    &\to -\frac{\Dmq_{12} \cos 2\theta_{12}}{4 E_\nu V(x)}
    + \frac{c_{13}^2}{2} \,,
    \\
    \big[ \xi^e + \xi^p + Y_n(x) \xi^n \big]
    \big(\chi^L + \chi^R \big)\, \Eps_N
    &\to -\frac{\Dmq_{12} \sin 2\theta_{12}}{4 E_\nu V(x)} \,,
  \end{aligned}
\end{equation}
where $V(x) \equiv \sqrt{2} G_F N_e(x)$ is the SM matter potential.
In other words, for a neutrino with a given energy $E_\nu$ at some
point $x$ along the trajectory, it is possible to find a pair of
values $(\Eps_D, \Eps_N)$ for which transitions are no longer
adiabatic.
Note that the cancellation condition for $\Eps_D$ depends on
$\cos2\theta_{12}$ and therefore will take different values for the
LMA and LMA-D regions, while the corresponding value for $\Eps_N$ will
remain invariant under a change of octant for $\theta_{12}$.  Also, as
the cancellation conditions in Eqs.~\eqref{eq:cancel} depend on the
values of $\xi^f$, the regions will depend on whether NSI take place
with electrons/protons, neutrons, or a combination of the two.

\begin{figure}\centering
  \includegraphics[width=\textwidth]{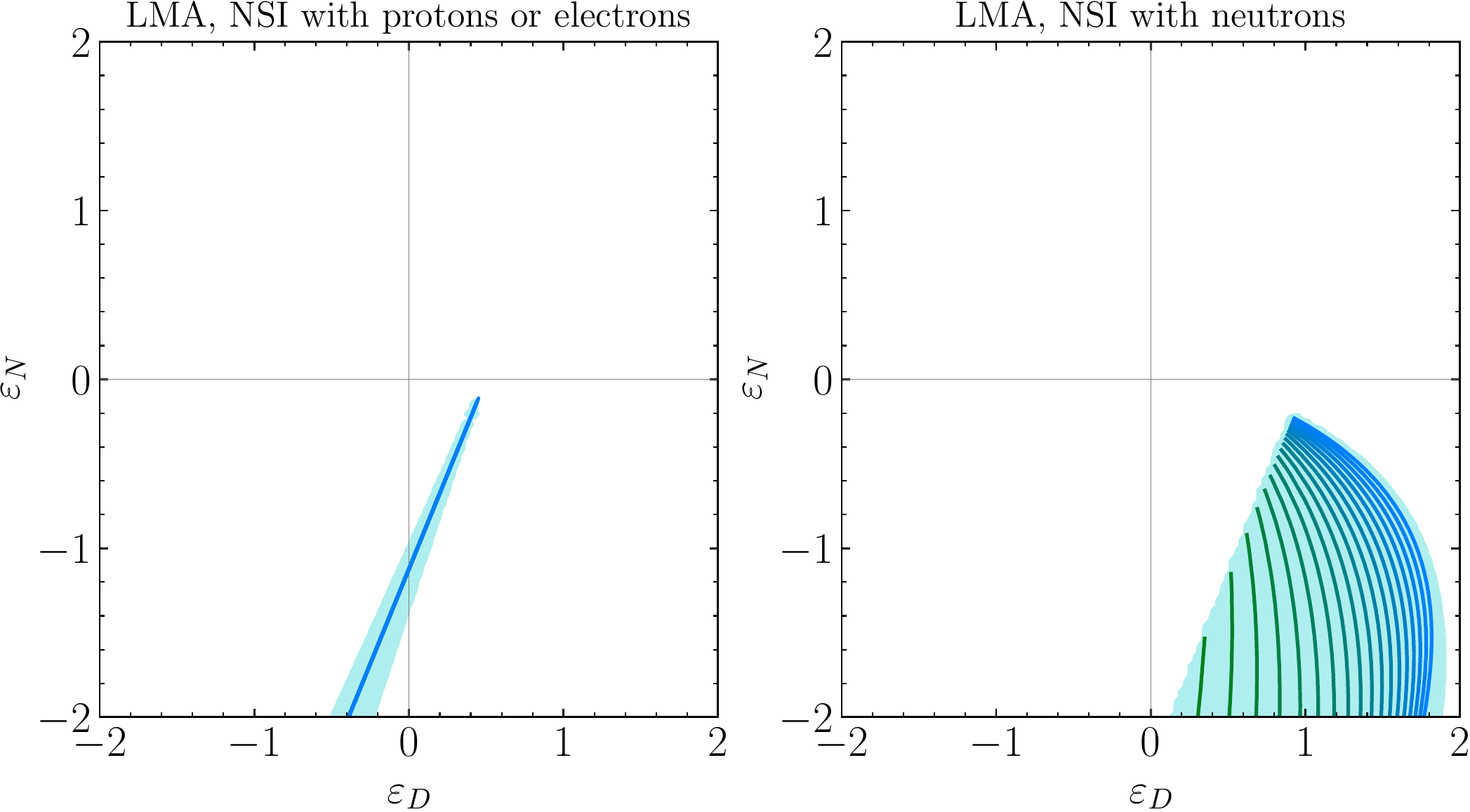}
  \caption{Departure from the adiabatic regime in presence of NSI, for
    a solar mixing angle in the LMA region.  The pale blue regions
    show the values of NSI parameters leading to $\gamma^{-1} > 1$,
    obtained from a numerical computation.  The coloured lines show
    the values of NSI parameters where Eqs.~\eqref{eq:cancel} are
    satisfied at some point along the neutrino trajectory, for a
    neutrino produced at the center of the Sun.  The different lines
    in the right panel correspond to neutrino energies between 1 MeV
    (green lines, at the left edge of the region) and 20~MeV (blue
    lines, towards the right edge).}
  \label{fig:adiab}
\end{figure}

Figure~\ref{fig:adiab} shows the regions where the transitions are not
adiabatic, for NSI with protons or electrons (left panel) and for NSI
with neutrons (right panel).  The shaded pale blue regions show the
results from a numerical computation.  In contrast, the coloured lines
show the points satisfying the analytic conditions in
Eqs.~\eqref{eq:cancel}, for neutrino energies between 1~MeV (green
lines, towards the left edge of the region) and 20~MeV (blue lines,
towards the right edge of the region), for $\theta_{12} = 33^\circ$
and $\Dmq_{12} = 7.5\times 10^{-5}~\eVq$.  As can be seen, the
agreement with the numerical computation is excellent.  Also, note the
very different shape of the regions in the two panels.  The reason
behind this is that for NSI with electrons or protons the dependence
with $E_\nu$ and $x$ comes in Eqs.~\eqref{eq:cancel} through the
product $E_\nu V(x)$, while for NSI with neutrons there is an extra
dependence on $Y_n(x)$.  Therefore, while the regions in the left
panel in Fig.~\ref{fig:adiab} span essentially a straight line, in the
right panel the dependence is more complex.

In our past study~\cite{Esteban:2018ppq}, where only NSI with quarks
were considered, neutrino evolution in the Sun was based on a fully
numerical approach.  In the present work, however, exploring the full
parameter space of the most general NSI with protons, electrons, and
neutrons without assuming adiabaticity is challenging from the
numerical point of view.  To overcome this problem in our calculations
we start by evaluating the adiabaticity index for the point in the NSI
and oscillation parameter space to be surveyed.  If for that point
$\gamma^{-1}(x) < 1$ along the whole neutrino trajectory in the Sun we
use the adiabatic approximation when computing the corresponding
flavour transition probabilities and the subsequent prediction for all
observables and $\chi^2$ value.  If, on the contrary, the adiabaticity
condition is violated somewhere inside the Sun such point is removed
from the parameter space to be surveyed.  We do so because we have
verified that for parameter values for which adiabaticity in the Sun
is violated the predicted observables with the properly computed
flavour transition probability without assuming adiabaticity never
lead to good description of the data.  But we also find that if one
evaluates the probabilities for those parameters wrongly using the
adiabatic approximation, one can find a ``fake'' good fit to the Solar
and KamLAND data, which would lead to wrong conclusions about their
acceptability.  In conclusion, the adiabatic approximation can be
safely used for the purposes of this work as long as one removes from
the parameter space those points for which adiabaticity in the Sun is
violated.  Furthermore, once the data of atmospheric neutrino
experiments is added to the fit, it totally disfavors the parameter
regions where transitions in the Sun are not adiabatic.

\subsection{Neutrino detection cross sections in the presence of NSI}
\label{sec:formCS}

In addition to propagation effects discussed above, non-standard
interactions can also affect scattering in the detector.  In this
respect, it should be mentioned that all the atmospheric, reactor and
accelerator experiments included in our fit rely on CC processes in
order to detect neutrinos, so the corresponding cross-section is not
affected by the NC-like NSI considered here.\footnote{It should be
noted that a subleading component of NC interactions is present as
background in many of these experiments, so that NSI parameter may in
principle have an impact on the number of events.  However, we expect
these effects to be very small, so for simplicity we neglect them.}
This implies that the $\sigma^\text{det}$ generalized cross-section
matrix entering Eq.~\eqref{eq:ES-dens} is diagonal in the SM flavour
basis, and its non-zero entries coincide with the usual SM
cross-section, that is,
\begin{equation}
  \sigma^\text{det} = \diag(
  \sigma_e^\text{SM},\, \sigma_\mu^\text{SM},\, \sigma_\tau^\text{SM}) \,
  \qquad\text{for CC processes.}
\end{equation}
Some solar neutrino experiments, on the contrary, are sensitive to NC
NSI in some of the detection processes involved.  This is the case of
Borexino and SK (and SNO, albeit with lower sensitivity) which observe
neutrino-electron ES, which may be affected by electron NSI.
Regarding NSI with nuclei SNO can also probe axial-vector NSI in NC
events, and CE$\nu$NS experiments are able to set important
constraints on vector interactions.  In the rest of this section we
review the phenomenological implications for these three cases
separately.

\subsubsection{Neutrino-electron elastic scattering}
\label{sec:formES}

The presence of flavour-changing effects in NSI implies that the SM
flavour basis no longer coincides with the interaction eigenstates of
the neutrino-electron scattering.  In such case the generalized cross
section $\sigma^\text{det}$ can be obtained as the integral over the
electron recoil kinetic energy $T_e$ of the following matrix
expression:
\begin{multline}
  \label{eq:nsi-elec}
  \dfrac{\dd\sigma^\text{ES}}{\dd T_e}(E_\nu, T_e)
  = \dfrac{2 G_F^2 m_e}{\pi}
  \bigg\lbrace
  C_L^2 \Big[ 1 + \dfrac{\alpha}{\pi} f_-(y) \Big]
  + C_R^2\, (1-y)^2 \Big[ 1 + \dfrac{\alpha}{\pi}f_+(y) \Big]
  \\
  - \big\{ C_L, C_R \big\}\,
  \dfrac{m_e y}{2E_\nu} \Big[ 1 + \dfrac{\alpha}{\pi} f_\pm(y) \Big]
  \bigg\rbrace
\end{multline}
where $y \equiv T_e / E_\nu$ and $f_+$, $f_-$, $f_\pm$ are loop
functions given in Ref.~\cite{Bahcall:1995mm}, while $\alpha$ stands
for the fine-structure constant and $m_e$ is the electron mass.  In
this formula $C_L$ and $C_R$ are $3\times 3$ hermitian matrices which
incorporate both SM and NSI contributions:
\begin{equation}
\label{eq:CL-CR}
  C_{\alpha\beta}^L
  \equiv c_{L\beta}\, \delta_{\alpha\beta} + \Eps_{\alpha\beta}^{e,L}
  \quad\text{and}\quad
  C_{\alpha\beta}^R
  \equiv c_{R\beta}\, \delta_{\alpha\beta} + \Eps_{\alpha\beta}^{e,R} \,.
\end{equation}
The effective couplings $c_{L\beta}$ and $c_{R\beta}$ account for the
SM part, and contain both the flavour-universal NC terms and the
$\nu_e$-only CC scattering:
\begin{equation}
  \label{eq:cm-coupl}
  \begin{aligned}
    c_{Le}
    &= \rho\,\Big[ \kappa_{e}(T_e)\sin^2\theta_w - \dfrac{1}{2} \Big] + 1 \,,
    &\quad
    c_{Re}
    &= \rho\, \kappa_{e}(T_e)\sin^2\theta_w \,,
    \\
    c_{L\tau} = c_{L\mu}
    &= \rho\, \Big[ \kappa_\mu(T_e)\sin^2\theta_w - \dfrac{1}{2} \Big] \,,
    &\quad
    c_{R\tau} = c_{R\mu}
    &= \rho\, \kappa_\mu(T_e)\sin^2\theta_w \,,
  \end{aligned}
\end{equation}
with $\theta_w$ being the weak mixing angle, and $\rho$ and
$\kappa_\beta(T_e)$ departing from $1$ due to radiative corrections of
the gauge boson self-energies and vertices~\cite{Bahcall:1995mm}.
It is immediate to see that, if the NSI terms
$\Eps_{\alpha\beta}^{e,L}$ and $\Eps_{\alpha\beta}^{e,R}$ are set to
zero, the matrix $\dd\sigma^\text{ES} \big/ \dd T_e$ becomes diagonal.
Imposing the factorization of Eq.~\eqref{eq:eps-fact} for the vector
($+$) and axial-vector ($-$) NSI we get:
\begin{equation}
  \Eps_{\alpha\beta}^{e,V(A)}=
  \Eps_{\alpha\beta}^{e,L}\pm \Eps_{\alpha\beta}^{e,R} =
    \Eps_{\alpha\beta} \, \xi^e\, (\chi^L \pm \chi^R)
    = \sqrt{5} \cos\eta \sin\zeta\,
    (\chi^L \pm \chi^R)\, \Eps_{\alpha\beta}\,.
\end{equation}

Let us finalize this section by discussing briefly the impact that the
inclusion of NSI effects on ES could have on the generalized
mass-ordering degeneracy discussed in Sec.~\ref{sec:formOSC}.  We have
seen that the parameter transformations~\eqref{eq:osc-deg}
and~\eqref{eq:NSI-deg} lead to a complex conjugation of the neutrino
density matrix, $\rho^\text{det} \to [\rho^\text{det}]^*$.  As shown
in Eq.~\eqref{eq:conjugate}, this does not affect the overall number
of events (thus resulting in the appearance of the degeneracy) as long
as it is accompanied by a similar transformation $\sigma^\text{det}
\to [\sigma^\text{det}]^*$.  The latter can be realized either as $C_L
\to C_L^*$ and $C_R \to C_R^*$, which occur when both
$\Eps_{\alpha\beta}^{e,L}$ and $\Eps_{\alpha\beta}^{e,R}$ undergo
simple complex conjugation, or as $C_L \to -C_L^*$ and $C_R \to
-C_R^*$, which require ad-hoc transformations of the diagonal entries
$\Eps_{\alpha\alpha}^{e,L}$ and $\Eps_{\alpha\alpha}^{e,R}$ to
compensate for the SM contribution of Eq.~\eqref{eq:cm-coupl}.  While
conceptually identical to the situation occurring in neutrino
oscillations, where the extra freedom introduced by NSI allows to
``flip the sign'' of the standard matter effects, the specific
transformations required to achieve a perfect symmetry of
$\sigma^\text{det}$ differ from those of Eq.~\eqref{eq:NSI-deg}.  In
principle one may first choose $\Eps_{\alpha\beta}^{e,L}$ and
$\Eps_{\alpha\beta}^{e,R}$ accounting for Eq.~\eqref{eq:cm-coupl} and
then tune $\Eps_{\alpha\beta}^{p,V}$ and $\Eps_{\alpha\beta}^{n,V}$ to
fulfill Eq.~\eqref{eq:NSI-deg}, but this procedure is incompatible
with the factorization constraint of Eq.~\eqref{eq:eps-fact}, which
assumes that all NSI have the same neutrino flavour structure
independently of their chirality and of the charged fermion type.  The
net conclusion is that, for NSI involving electrons (that is,
$\zeta\ne 0$) and assuming that the factorization in
Eq.~\eqref{eq:eps-fact} holds, ES effects \emph{break} the generalized
mass-ordering degeneracy.

\subsubsection{SNO neutral-current cross-section}
\label{sec:sno-nc}

The SNO experiment observed NC interactions of solar neutrinos on
deuterium.  At low energies, the corresponding cross section is
dominated by the Gamow-Teller transition and it scales as $g_A^2$
where $g_A$ is the coupling of the neutrino current to the axial
isovector hadronic current which in the SM is given by $g_A\equiv
g_A^u-g_A^d$~\cite{Bahcall:1988em, Bernabeu:1991sd, Chen:2002pv}.
Using that the nuclear corrections to $g_A$ are the same when the NSI
are added, we obtain that in the presence of the NC NSI we can write
\begin{equation}
\label{eq:sigma_axial}
  \sigma^\text{det} = \sigma_\text{SM} \bigg(\frac{G_A}{g_A}\bigg)^2
\end{equation}
where $G_A$ is an hermitian matrix in flavour space
\begin{equation}
  \frac{G_A}{g_A}
  = \delta_{\alpha\beta}
  + \Eps_{\alpha\beta}^{u,A} -\Eps_{\alpha\beta}^{d,A}
  = \delta_{\alpha\beta} +
  \Eps_{\alpha\beta} \, (\xi^u - \xi^d) \, (\chi^L - \chi^R) \,,
\end{equation}
Clearly for vector NSI ($\chi^L =\chi^R$), the NSI contributions
vanish and $\sigma^\text{det}$ takes just the SM value times the
identity in flavour space.  Conversely for axial-vector NSI one gets
$\chi^L - \chi^R = 1$ and the NSI term contributes.

\subsubsection{Coherent elastic neutrino-nucleus scattering}
\label{sec:formCNUES}

The generalized cross-section $\sigma^\text{det}$ describing CE$\nu$NS
in the presence of NSI can be obtained by integrating over the recoil
energy of the nucleus $E_R$ the following expression:
\begin{equation}
  \label{eq:xsec-SM}
  \frac{\dd\sigma^\text{coh} (E_R, E_\nu)}{\dd E_R}
  = \frac{G_F^2}{2\pi} \,
  \mathcal{Q}^2 \, F^2(q^2) \, m_A
  \bigg(2 - \frac{m_A E_R}{E_\nu^2} \bigg)
\end{equation}
where $m_A$ is the mass of the nucleus and $F(q^2)$ is its nuclear
form factor evaluated at the squared momentum transfer of the process,
$q^2 = 2 m_A E_R$.  In this formalism, the structure in flavour space
which characterizes $\sigma^\text{det}$ is encoded into the hermitian
matrix $\mathcal{Q}$, which is just the generalization of the weak
charge of the nucleus for this formalism.  For a nucleus with $Z$
protons and $N$ neutrons, it reads:
\begin{equation}
  \label{eq:Qweak}
  \mathcal{Q}_{\alpha\beta}
  = Z \big(g_p^V \delta_{\alpha\beta} + \Eps_{\alpha\beta}^{p,V} \big)
  + N \big(g_n^V \delta_{\alpha\beta} + \Eps_{\alpha\beta}^{n,V} \big)
\end{equation}
where $g_p^V = 1/2 - 2\sin^2\theta_w$ and $g_n^V = -1/2$ are the SM
vector couplings to protons and neutrons, respectively.  In
experiments with very short baselines such as those performed so far,
neutrinos have no time to oscillate and therefore the density matrix
at the detector $\rho^\text{det}$ is just the identity matrix.  Taking
this explicitly into account in Eq.~\eqref{eq:ES-dens} we get:
\begin{equation}
  \rho^\text{det} = I
  \quad\Rightarrow\quad
  N_\text{ev} \propto \mathcal{Q}_\alpha^2
  \quad\text{with}\quad
  \mathcal{Q}_\alpha^2 \equiv \big[ \mathcal{Q}^2 \big]_{\alpha\alpha}
  = (\mathcal{Q}_{\alpha\alpha})^2
  + \sum_{\beta\ne\alpha} |\mathcal{Q}_{\alpha\beta}|^2
\end{equation}
for incident neutrino flavour $\alpha$, thus recovering the
expressions for the ordinary weak charges $\mathcal{Q}_\alpha^2$ used
in our former publications.  Coming back to Eq.~\eqref{eq:Qweak}, let
us notice that it can be rewritten as:
\begin{equation}
  \mathcal{Q}_{\alpha\beta}
  = Z \big[ (g_p^V + Y_n^\text{coh} g_n^V)\, \delta_{\alpha\beta}
  + \Eps_{\alpha\beta}^\text{coh} \big]
  \quad\text{with}\quad
  \Eps_{\alpha\beta}^\text{coh}
  \equiv \Eps_{\alpha\beta}^{p,V} + Y_n^\text{coh} \Eps_{\alpha\beta}^{n,V}
\end{equation}
where $Y_n^\text{coh} \equiv N/Z$ is the neutron/proton ratio
characterizing the target of a given CE$\nu$NS experiment.  Imposing
the quark-lepton factorization of Eq.~\eqref{eq:eps-fact} we get:
\begin{equation}
  \Eps_{\alpha\beta}^\text{coh}
  = \Eps_{\alpha\beta} \big( \xi^p +Y_n^\text{coh} \xi^n \big)
  \big( \chi^L + \chi^R \big)
  = \sqrt{5} \, \big[\! \cos\eta\, \cos\zeta
    + Y_n^\text{coh} \sin\eta \big] \big( \chi^L + \chi^R \big)\,
  \Eps_{\alpha\beta}\,.
\end{equation}
Similarly to Eq.~\eqref{eq:eps-earth}, this expression suggests that
the analysis of coherent scattering data can be performed in terms of
the effective couplings $\Eps_{\alpha\beta}^\text{coh}$, whose
best-fit value and allowed ranges are independent of $(\eta, \zeta)$.
As a consequence, the bounds on $\Eps_{\alpha\beta}$ simply scale as
$[\cos\eta \cos\zeta + Y_n^\text{coh} \sin\eta]$.  In analogy to
Eq.~\eqref{eq:epx-etapr}, one can define an effective angle
$\tan\eta^{\prime\prime} \equiv \tan\eta\, \big/ \cos\zeta$
parametrizing the direction in the $(\xi^p, \xi^n)$ plane, such that
all CE$\nu$NS experiments depend on $(\eta, \zeta)$ only thought the
combination $\eta^{\prime\prime}$.  In particular, it is
straightforward to see that a coherent scattering experiment
characterized by a given $Y_n^\text{coh}$ ratio will yield no bound on
$\Eps_{\alpha\beta}$ for $\eta^{\prime\prime} =
\arctan(-1/Y_n^\text{coh})$, as for this value the effects of NSI on
protons and neutrons cancel exactly.
Such a cancellation can be obtained, for example, for models of the
type proposed in Ref.~\cite{Bernal:2022qba}, where the $Z'$ associated
to a new gauge symmetry also has a sizable kinetic mixing with the SM
photon, which allows for arbitrary relative size of NSI with up and
down quarks.

\section{Results}
\label{sec:results}

This section summarizes the main results of our study.  In
Sec.~\ref{sec:sim} we first review the data included in the fit,
introduce our $\chi^2$ definition, and outline the details related to
the sampling of the multi-dimensional parameter space.  We then
proceed to present our results for NSI with electrons
(Sec.~\ref{sec:resule}), with quarks (Sec.~\ref{sec:resulq}), and for
simultaneous NSI with electrons and quarks (Sec.~\ref{sec:resulgen}).
The status of the LMA-D solution in this general case is discussed in
Sec.~\ref{sec:resulLMAD}.

\subsection{Simulation details}
\label{sec:sim}

The data samples included in our oscillation analysis mostly coincide
with those in NuFIT-5.2~\cite{nufit-5.2}.  In brief, in the analysis
of solar neutrino data we consider the total rates from the
radiochemical experiments Chlorine~\cite{Cleveland:1998nv},
Gallex/GNO~\cite{Kaether:2010ag}, and SAGE~\cite{Abdurashitov:2009tn},
the spectral data (including day-night information) from the four
phases of Super-Kamkoikande in Refs.~\cite{Hosaka:2005um,
  Cravens:2008aa, Abe:2010hy,SK:nu2020}, the results of the three
phases of SNO in the form of the day-night spectrum data of
SNO-I~\cite{Aharmim:2007nv}, and SNO-II~\cite{Aharmim:2005gt} and the
three total rates of SNO-III~\cite{Aharmim:2008kc},\footnote{This
corresponds to the analysis labeled \textsc{SNO-data} in
Ref.~\cite{Gonzalez-Garcia:2013usa}.  In that article an alternative
analysis of the SNO data was introduced, labeled \textsc{SNO-poly},
based on an effective \emph{MSW-like} polynomial parametrization for
the day and night survival probabilities of the combined SNO phases
I--III, as detailed in Ref.~\cite{Aharmim:2011vm}.  The
\textsc{SNO-poly} approach can be efficiently used as long as NSI only
enter in propagation through the matter potential, and therefore in
the subsequent analysis~\cite{Esteban:2018ppq, Coloma:2019mbs} the SNO
results were included in this way.  However, the polynomial
parametrization of \textsc{SNO-poly} cannot account for NSI effects in
neutrino interactions, so that in the present work we have reverted to
the full \textsc{SNO-data} version of the SNO analysis.  This leads to
small differences in the results of the analysis with couplings to
quarks only with respect to those in Refs.~\cite{Esteban:2018ppq,
  Coloma:2019mbs}.} and the spectra from Borexino
Phase-I~\cite{Bellini:2011rx, Bellini:2008mr}, and
Phase-II~\cite{Borexino:2017rsf}.  For reactor neutrinos we include
the separate DS1, DS2, DS3 spectra from KamLAND~\cite{Gando:2013nba}
with Daya Bay reactor $\nu$ fluxes~\cite{An:2016srz}, the FD/ND
spectral ratio, with 1276-day (FD), 587-day (ND) exposures of
Double-Chooz~\cite{DoubleC:nu2020}, the 3158-day separate EH1, EH2,
EH3 spectra~\cite{DayaBay:2022orm} of Daya-Bay, and the 2908-day FD/ND
spectral ratio of RENO~\cite{RENO:nu2020}.  For atmospheric neutrinos
we use the four phases of Super-Kamiokande (up to 1775 days of
SK4~\cite{Wendell:2014dka}), the complete set of DeepCore 3-year
$\mu$-like events presented in Ref.~\cite{Aartsen:2014yll} and
publicly released in Ref.~\cite{deepcore:2016}, and the results on
$\nu_\mu$-induced upgoing muons reported by
IceCube~\cite{TheIceCube:2016oqi} based on one year of data taking.
Finally, for LBL experiments we include the final neutrino and
antineutrino spectral data on $\nu_e$-appearance and
$\nu_\mu$-disappearance in MINOS~\cite{Adamson:2013whj}, the
$19.7\times 10^{20}$ pot $\nu_\mu$-disappearance and $16.3\times
10^{20}$ pot $\bar\nu_\mu$-disappearance data in
T2K~\cite{T2K:nu2020}, and the $13.6\times 10^{20}$ pot
$\nu_\mu$-disappearance and $12.5\times 10^{20}$ pot
$\bar{\nu}_\mu$-disappearance data in NO$\nu$A~\cite{NOvA:nu2020}.
Notice that to ensure full consistency with our CP-conserving
parametrization we have chosen not to include in the present study the
data from the $\nu_e$ and $\bar\nu_e$ appearance channels in NO$\nu$A
and T2K.  With this data we construct
$\chi^2_\text{OSC}(\vec\omega,\vec\Eps)$ where we denote by
$\vec\omega$ the 3$\nu$ oscillation parameters and $\vec\Eps$ the NSI
parameters considered in the analysis.

When combining with CE$\nu$NS we include the results from COHERENT
data on CsI~\cite{COHERENT:2017ipa, COHERENT:2018imc} and Ar
targets~\cite{COHERENT:2020iec, COHERENT:2020ybo} (see
Refs.~\cite{Coloma:2019mbs, Coloma:2022avw} for details).  In
particular, for the analysis of COHERENT CsI data we use the quenching
factor from Ref.~\cite{Collar:2019ihs} and the nuclear form factor
from Ref.~\cite{Klos:2013rwa}.  For the analysis of COHERENT Ar data
we use the quenching factor provided by the COHERENT collaboration in
Ref.~\cite{COHERENT:2020ybo} and Helm~\cite{Helm:1956zz} nuclear form
factor (the values of the parameters employed are the same as in
Ref.~\cite{Coloma:2022avw}).  For CE$\nu$NS searches using reactor
neutrinos at Dresden-II reactor experiment~\cite{Colaresi:2022obx} we
follow the analysis presented in Ref.~\cite{Coloma:2022avw} with YBe
quenching factor~\cite{Collar:2021fcl} (the characteristic
momentum-transfer in this experiment is very low so the nuclear form
factor can be taken to be 1).  With all this we construct the
corresponding $\chi^2_\text{COH,CsI}(\vec\Eps)$,
$\chi^2_\text{COH,Ar}(\vec\Eps)$, and
$\chi^2_\text{D-II,Ge}(\vec\Eps)$.

Our goal is to find the global minimum of the total $\chi^2$ which,
unless otherwise stated, is obtained adding the contributions from our
global analysis of oscillation data (<<GLOB-OSC w NSI in ES>>) and
CE$\nu$NS data:
\begin{equation}
  \chi^2 (\vec\omega, \vec\Eps)
  = \chi^2_\text{OSC}(\vec\omega, \vec \Eps)
  + \chi^2_\text{CE$\nu$NS} (\vec\Eps)
\end{equation}
where we have defined
\begin{equation}
  \chi^2_\text{CE$\nu$NS}
  = \chi^2_\text{COH,CsI} + \chi^2_\text{COH,Ar} + \chi^2_\text{D-II,Ge} \,.
\end{equation}
The minimum of the total $\chi^2$ is obtained after minimization over
all the nuisance parameters, which are included as pull terms in our
fit.\footnote{For details on the numerical implementation of
systematic uncertainties see Refs.~\cite{Coloma:2022avw,
  Coloma:2022umy, nufit-5.2}.}  However, note that a priori it is
possible to suppress the effects of vector NSI in neutrino detection
while still allowing for sizable NSI in oscillations, since the
momentum transfer required for the effects to be observable is not the
same in the two cases~\cite{Farzan:2015doa}.  The paradigmatic example
of BSM scenarios leading to sizable NSI are models with light
mediators.  Matter effects arise from a coherent effect, that is, with
zero momentum-transfer, and therefore apply to NSI induced by ultra
light mediators, as long as their mass is $\Mmed \gtrsim 1/R_\oplus
\sim \mathcal{O}(10^{-12})$~eV~\cite{Gonzalez-Garcia:2006vp,
  Coloma:2020gfv}.  Scattering effects, on the contrary, require a
minimum momentum-transfer to be observable:
\begin{itemize}
\item In the case of ES, for recoil energies $T_e \simeq
  \mathcal{O}(500~\text{keV})$ (which are in the right ballpark for
  Borexino) the relevant scale is $q = \sqrt{2 m_e T_e} \sim
  \mathcal{O}(500~\text{keV})$.  Therefore, in what follows we will
  also present our results neglecting the effects due to NSI in
  detection (<<GLOB-OSC w/o NSI in ES>>), so as to consider this class
  of models with very light mediators, $ \Mmed \ll
  \mathcal{O}(500~\text{keV})$.  For SNO and SK, sensitive to
  neutrinos with higher energies, a similar argument leads to $q \sim
  \mathcal{O}(5-10~\text{MeV})$.  Hence, results labeled as <<GLOB-OSC
  w NSI in ES>> apply to NSI generated by models with $\Mmed \gtrsim
  10~\text{MeV}$, where we expect to have NSI effects on the detection
  through ES for all solar experiments considered in this work.
  Finally, in the intermediate mass range only Borexino would be fully
  sensitive to NSI effects on detection.

\item In the case of CE$\nu$NS, the relevant momentum transfer also
  depends on the experiment considered.  For COHERENT data on CsI
  (Ar), this can be estimated as $q \sim 50~\text{MeV}$ ($q \sim
  30~\text{MeV}$) while for Dresden-II we get $q \sim
  \mathcal{O}(5~\text{MeV})$.  Therefore, in Sec.~\ref{sec:resulq} we
  will present our results using the combination of all oscillation
  and CE$\nu$NS data, sensitive to models with $\Mmed \gtrsim
  50~\text{MeV}$ (<<GLOB-OSC+CE$\nu$NS>>), as well as the results for
  oscillation data alone, sensitive to NSI models with $\Mmed\ll
  5~\text{MeV}$ (labeled as <<GLOB-OSC>>).  In the intermediate mass
  range between 5 and 50 MeV, effects on CE$\nu$NS may be suppressed
  for some experiments but not all of them.
\end{itemize}
The intermediate mass ranges defined above (where NSI effects in
detection may be observed at one experiment, but not another) require
special handling and are therefore beyond the scope of the present
paper, leaving them for future work.

In the axial-vector case there are no NSI effects in oscillations nor
CE$\nu$NS and the NSI parameters only enter in the ES and/or NC
detection cross sections.  Thus for axial-vector NSI with electrons
the derived bounds apply for mediator masses as discussed above for
ES.  NC interactions in SNO occur for neutrinos with energy above the
binding energy of the deuteron, 2.22~MeV.  So the bounds for
axial-vector NSI with quarks derived here apply for $\Mmed \gtrsim
3~\text{MeV}$.

Let us finalize this discussion by briefly reviewing the number of
parameters involved in the fit.  A priori, under the approximations
assumed in the present work, atmospheric and long-baseline experiments
depend on all the oscillation parameters as well as on the NSI
variables defined in Eqs.~\eqref{eq:eps_atm}.  However, for LBL
experiments we fix $\Dmq_{21}$ and $|\theta_{12} - 45^\circ|$ to their
SM best-fit value, as their relevance for such data is only subleading
and we know from previous studies~\cite{Esteban:2018ppq} that NSI are
not going to spoil their determination sizably (except for the
$\theta_{12}$ octant which we do \emph{not} fix, hence the absolute
value in the corresponding fixed value above).  For the analysis of
atmospheric neutrinos we further set $\Dmq_{21} \to 0$, so the
dependence on $\theta_{12}$ and $\delta_\text{CP}$ vanishes.  With
this, neutrino propagation through the Earth is described in terms of
six continuous parameters, namely ($\Dmq_{31}$, $\theta_{13}$,
$\theta_{23}$) for the vacuum part and ($\Eps_\oplus$, $\varphi_{12}$,
$\varphi_{13}$) for the matter part, as well as the two CP-conserving
values of $\delta_\text{CP}$ and the two options for the $\theta_{12}$
octant.  We additionally fix the value of $\theta_{13}$ to the present
best fit, as its determination comes mainly from reactor experiments
for which NSI effects are suppressed.  On the other hand, neutrino
oscillations in solar and KamLAND data depend effectively on $\Eps_D$
and $\Eps_N$ for the matter part, while the vacuum part depends on
$\theta_{12}$, $\theta_{23}$, $\Dmq_{21}$ and the two-CP conserving
values of $\delta_\text{CP}$ ($\theta_{13}$ is also fixed to its
best-fit value here).  Finally note that, while oscillations depend on
differences between the diagonal NSI parameters, the addition of
scattering effects in detection breaks this pattern.  This implies
that in addition to the two differences in Eqs.~\eqref{eq:eps-earth}
we need to scan one of the diagonal parameters, which we take to be
$\Eps_{\mu\mu}^\oplus$.  The last two parameters of relevance are the
two angles $\eta$ and $\zeta$, which determine the direction of NSI in
$(\xi^e,\xi^p,\xi^n)$ space.

Taking into account all of the above, in our simulations we scan the
multi-dimensional space spanned by the following continuous
parameters:
\begin{equation}
  \label{eq:10param}
  \theta_{12},\, \theta_{23},\, \Dmq_{21},\, \Dmq_{31},\,
  \Eps_\oplus,\, \varphi_{12},\, \varphi_{23},\,
  \Eps_{\mu\mu}^\oplus,\, \eta,\, \zeta,
\end{equation}
as well as the two CP-conserving values of $\delta_\text{CP} \in
\lbrace 0, \pi \rbrace$.  In order to scan this multi-dimensional
parameter space efficiently, we make use of the
MultiNest~\cite{Feroz:2013hea, Feroz:2008xx} and
Diver~\cite{Martinez:2017lzg} algorithms.

\begin{figure}\centering
  \includegraphics[width=\textwidth]{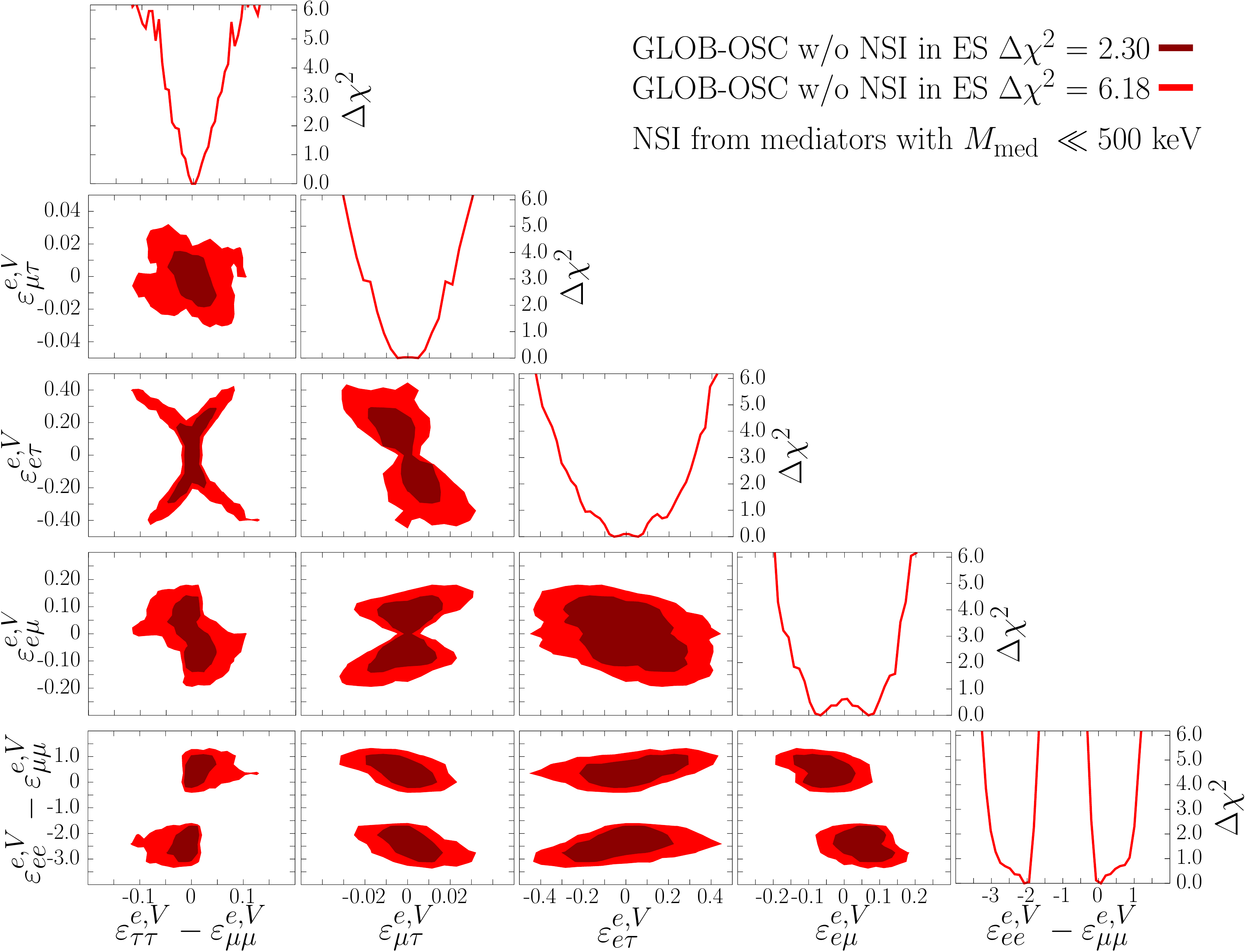}
  \caption{Constraints on the coefficients for vector NSI with
    electrons from the global analysis of oscillation data
    \textit{without including the effect of NSI in the detection cross
      section}.  Each panel shows a two-dimensional projection of the
    allowed multi-dimensional parameter space after minimization with
    respect to the undisplayed parameters.  The regions correspond to
    $1\sigma$ and $2\sigma$ (2 d.o.f.).}
  \label{fig:etriangVwoES}
\end{figure}

\begin{table}\centering
  \catcode`?=\active\def?{\hphantom{0}}
  \begin{tabular}{|l||c|c|}
    \hline
    & \multicolumn{2}{|c|}{Allowed ranges at 90\% CL (marginalized)}
    \\
    \hline
    & \multicolumn{2}{|c|}{GLOB-OSC w/o NSI in ES}
    \\
    \hline
    & LMA & $\text{LMA}\oplus\text{LMA-D}$
    \\
    \hline
    $\Eps_{ee}^{e,V} - \Eps_{\mu\mu}^{e,V}$
    & $[-0.21, +1.0]?$
    & $[-3.0, -1.8] \oplus[-0.21, +1.0]$
    \\
    $\Eps_{\tau\tau}^{e,V} - \Eps_{\mu\mu}^{e,V}$
    & $[-0.015, +0.048]$
    & $[-0.040, +0.047]$
    \\
    $\Eps_{e\mu}^{e,V}$
    & $?[-0.15, +0.035]$
    & $[-0.15, +0.14]$
    \\
    $\Eps_{e\tau}^{e,V}$
    & $[-0.21, +0.31]$
    & $[-0.29, +0.31]$
    \\
    $\Eps_{\mu\tau}^{e,V}$
    & $[-0.020, +0.012]$
    & $[-0.020, +0.017]$
    \\
    \hline
    \end{tabular}
  \caption{90\% CL bounds (1 d.o.f., 2-sided) on the coefficients of
    NSI operators with electrons after marginalizing over all other
    NSI and oscillation parameters.  The bounds are derived from the
    global analysis of oscillation data \textit{without including the
      effect of NSI in the ES cross section} (NSI induced by mediators
    with mass $\Mmed \ll 500~\text{keV}$, see Sec.~\ref{sec:sim}).
    The ranges in the first column (labeled <<LMA>>) correspond to an
    analysis in which we restrict $\theta_{12} <45^\circ$.  In the
    second column (labeled <<$\text{LMA}\oplus\text{LMA-D}$>>), both
    $\theta_{12} <45^\circ$ and $\theta_{12} >45^\circ$ are allowed.
    The same bounds hold for vector NSI with protons.}
  \label{tab:nsieVwoES}
\end{table}

\subsection{New constraints on NSI with electrons}
\label{sec:resule}

\begin{figure}\centering
  \includegraphics[width=\textwidth]{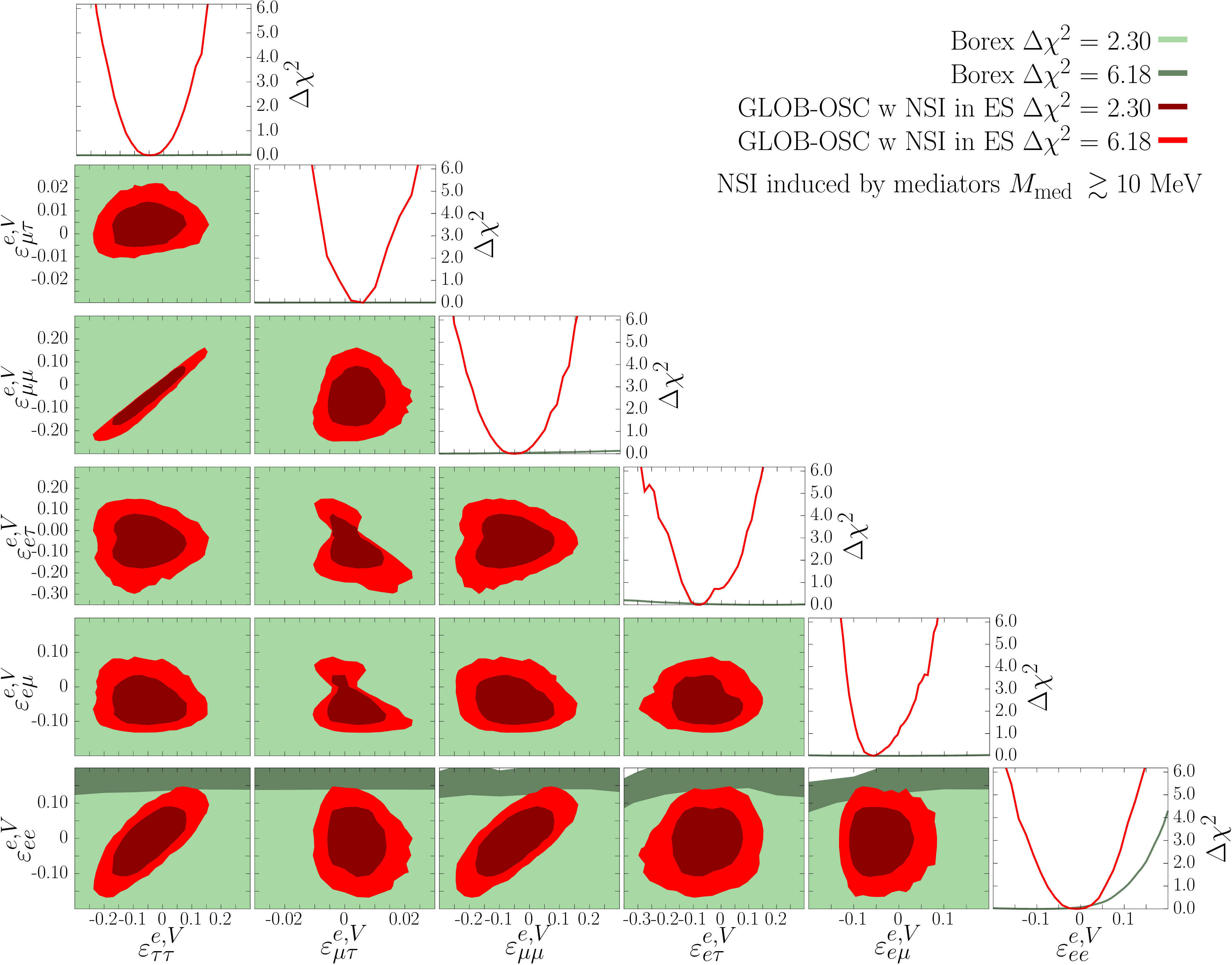}
  \caption{Constraints on the coefficients for vector NSI with
    electrons from the global analysis of oscillation data
    \emph{including the effect of NSI in the ES cross section}.  Each
    panel shows a two-dimensional projection of the allowed
    multi-dimensional parameter space after minimization with respect
    to the undisplayed parameters.  The contours correspond to
    $1\sigma$ and $2\sigma$ (2 d.o.f.).  The closed red regions
    correspond to the global oscillation analysis which involves the
    six NSI plus five oscillation parameters.  For the sake of
    comparison we also show as green regions the constraints obtained
    from the analysis of full Borexino Phase-II spectrum in
    Ref.~\recite{Coloma:2022umy}.}
  \label{fig:etriangVwES}
\end{figure}

We start by describing the results of the global analysis for the case
of NSI with electrons, which, as mentioned in the introduction, was
not considered in our previous studies~\cite{Gonzalez-Garcia:2013usa,
  Esteban:2018ppq, Coloma:2019mbs}.  We plot in
Fig.~\ref{fig:etriangVwoES} and~\ref{fig:etriangVwES} the constraints
on the different coefficients for the two scenarios outlined in the
previous section, without and with NSI effects in the detection cross
section, respectively.\footnote{Notice that from the point of view of
the data analysis the results from the <<GLOB-OSC w/o NSI in ES>> are
totally equivalent to those obtained for vector NSI which couple only
to protons.}  The corresponding 90\% CL allowed ranges are listed in
Table~\ref{tab:nsieVwoES} and on the left columns in
Table~\ref{tab:nsiewES} respectively.

\begin{table}\centering
  \catcode`?=\active\def?{\hphantom{0}}
  \renewcommand{\arraystretch}{1.2}
  \begin{tabular}{|c || c | c || c | c ||}
    \hline
    & \multicolumn{4}{c||}{Allowed ranges at 90\% CL (marginalized)}
    \\
    \hline
    & \multicolumn{2}{c||}{Vector ($X=V$)}
    & \multicolumn{2}{c||}{Axial-vector ($X=A$)}
    \\
    \hline
    & Borexino & \small{GLOB-OSC w NSI in ES}
    & Borexino & \small{GLOB-OSC w NSI in ES}
    \\
    \hline
    $\Eps^{e,X}_{ee}$
    & $?[-1.1, +0.17]$
    & $[-0.13, +0.10]$
    & $[-0.38, +0.24]$
    & $[-0.13, +0.11]$
    \\
    $\Eps^{e,X}_{\mu\mu}$
    & $[-2.4, +1.5]$
    & $[-0.20, +0.10]$
    & $[-1.5, +2.4]$
    & $[-0.70, +1.2]?$
    \\ $\Eps^{e,X}_{\tau\tau}$
    & $[-2.8, +2.1]$
    & $?[-0.17, +0.093]$
    & $[-1.8, +2.8]$
    & $[-0.53, +1.0]?$
    \\
    $\Eps^{e,X}_{e\mu}$
    & $[-0.83, +0.84]$
    & $[-0.097, +0.011]$
    & $[-0.79, +0.76]$
    & $[-0.41, +0.40]$
    \\
    $\Eps^{e,X}_{e\tau}$
    & $[-0.90, +0.85]$
    & $?[-0.18, +0.080]$
    & $[-0.81, +0.78]$
    & $[-0.36, +0.36]$
    \\
    $\Eps^{e,X}_{\mu\tau}$
    & $[-2.1, +2.1]$
    & $[-0.0063, +0.016]?$
    & $[-1.9, +1.9]$
    & $[-0.79, +0.81]$
    \\
    \hline
  \end{tabular}
  \caption{90\% CL bounds (1 d.o.f., 2-sided) on the coefficients of
    vector NSI operators with electrons after marginalizing over all
    other NSI and oscillation parameters.  The bounds are derived from
    the global analysis of oscillation data \textit{including the
      effect of NSI in the ES cross section}.  For comparison, the
    results obtained from the analysis of Borexino Phase-II data in
    Ref.~\cite{Coloma:2022umy} are also shown for comparison.  Note
    that these bounds apply to interactions induced by mediators with
    masses $\Mmed \gtrsim 10~\text{MeV}$, see Sec.~\ref{sec:sim}.}
  \label{tab:nsiewES}
\end{table}

The first thing to notice is that in the scenario <<GLOB-OSC w/o NSI
in ES>>, for the flavour diagonal coefficients, only the combinations
$\Eps_{ee}^{e,V} - \Eps_{\mu\mu}^{e,V}$ and $\Eps_{\tau\tau}^{e,V} -
\Eps_{\mu\mu}^{e,V}$ can be constrained, and two separate allowed
ranges appear (see bottom row in Fig.~\ref{fig:etriangVwoES}): one
around $\Eps_{ee}^{e,V} - \Eps_{\mu\mu}^{e,V}\sim 0$ and another
around $\Eps_{ee}^{e,V} - \Eps_{\mu\mu}^{e,V}\sim -2$.  This is
nothing else than the result of the generalized mass-ordering
degeneracy of Eq.~\eqref{eq:NSI-deg}.  The two disjoint allowed
solutions correspond to regions of oscillation parameters with
$\theta_{12} <45^\circ$ (with ranges labeled <<LMA>> in
Table~\ref{tab:nsieVwoES}) and with $\theta_{12} >45^\circ$ (with
ranges labeled <<LMA-D>> in Table~\ref{tab:nsieVwoES}) respectively.
As discussed in Sec.~\ref{sec:formOSC}, the degeneracy is partly
broken by the variation of chemical composition of the matter along
the neutrino trajectory when NSI coupling to neutrons are involved.
But in this case, with only coupling to electrons, the degeneracy is
perfect as can be seen by the fact that $\Delta\chi^2=0$ in both
minima of $\Eps_{ee}^{e,V} - \Eps_{\mu\mu}^{e,V}$.
From the panels on the lower row we observe that the projection of the
allowed regions corresponding to each of the two solutions partly
overlap for $\Eps_{\tau\tau}^{e,V} - \Eps_{\mu\mu}^{e,V}$ and
$\Eps_{\alpha\neq\beta}^{e,V}$, which leads to the non-trivial shapes
of some of the corresponding two-dimensional regions for some pairs of
those paremeters.

\begin{figure}\centering
  \includegraphics[width=\textwidth]{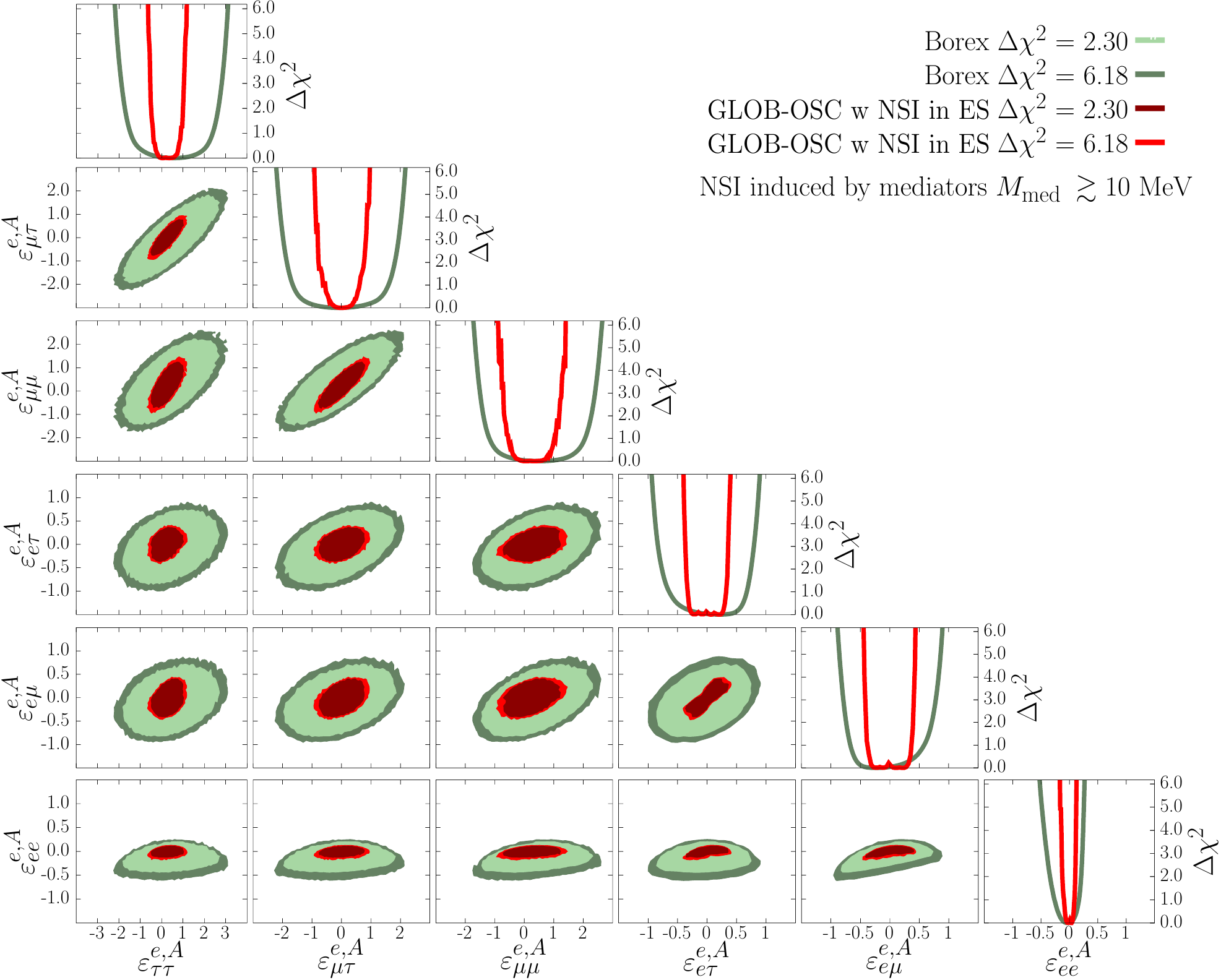}
  \caption{Same as Fig.~\ref{fig:etriangVwES} but for Axial-vector
    NSI.}
  \label{fig:etriangA}
\end{figure}

The results for the scenario <<GLOB-OSC w NSI in ES>> in
Fig.~\ref{fig:etriangVwoES}, and on the left columns in
Table~\ref{tab:nsiewES}, show that the inclusion of the effect of the
vector NSI in the ES cross sections in Borexino, SNO, and SK totally
lifts the degeneracy.  In this scenario only the LMA solution is
allowed, and $\Eps_{ee}^{e,V}$, $\Eps_{\mu\mu}^{e,V}$, and
$\Eps_{\tau\tau}^{e,V}$ can be independently constrained.  Notice also
that in this case, the lifting of the LMA-D solution leads to allowed
regions with more standard (close-to-elliptical) shapes.  For the sake
of comparison we show for this scenario the corresponding bounds
derived from the analysis of Borexino spectra in
Ref.~\cite{Coloma:2022umy}.  The comparison shows that for vector NSI
with electrons the global analysis of the oscillation data reduces the
allowed ranges of the NSI coefficients by factors $\sim\text{4--200}$
with respect to those derived with Borexino spectrum only.  In other
words, the NSI contribution to ES is important to break the LMA-D
degeneracy and to impose independent bounds on the three
flavour-diagonal NSI coefficients, but within the LMA solution the
effect of the vector NSI on the matter potential also leads to
stronger constraints.  This is further illustrated by the results
obtained for the analysis with axial-vector NSI which are shown in
Fig.~\ref{fig:etriangA} and right columns in Table~\ref{tab:nsiewES}.
Axial-vector NSI do not contribute to the matter potential and
therefore the difference between the results of the global oscillation
and the Borexino-only analysis in this case arises solely from the
effect of the axial-vector NSI on the ES cross section in SNO and SK.
Comparing the red and green regions in Fig.~\ref{fig:etriangA} and the
two left columns in Table~\ref{tab:nsiewES} we see that for
axial-vector NSI the improvement over the bounds derived with
Borexino-only analysis is just a factor $\sim$ 2--3.

\subsection{Updated constraints on NSI with quarks}
\label{sec:resulq}

Next we briefly summarize the results of the analysis for the
scenarios of NSI with either up or down quarks (more general
combinations of couplings to quarks and electrons will be presented in
the next section).  For vector NSI this updates and complements the
results presented in Refs.~\cite{Esteban:2018ppq, Coloma:2019mbs} by
accounting for the effects of increased statistics in the oscillation
experiments, including the addition of new data from Borexino
Phase-II.  Notice also that in the present analysis, as mentioned
above, the treatment of the SNO data is different than in
Refs.~\cite{Esteban:2018ppq, Coloma:2019mbs}.  Furthermore when
combining with CE$\nu$NS we include here the results from COHERENT
both on CsI~\cite{COHERENT:2017ipa, COHERENT:2018imc} and Ar
targets~\cite{COHERENT:2020iec, COHERENT:2020ybo}, together with the
recent results from CE$\nu$NS searches using reactor neutrinos at
Dresden-II reactor experiment~\cite{Colaresi:2022obx, Coloma:2022avw}.

\begin{figure}\centering
  \includegraphics[width=0.7\textwidth]{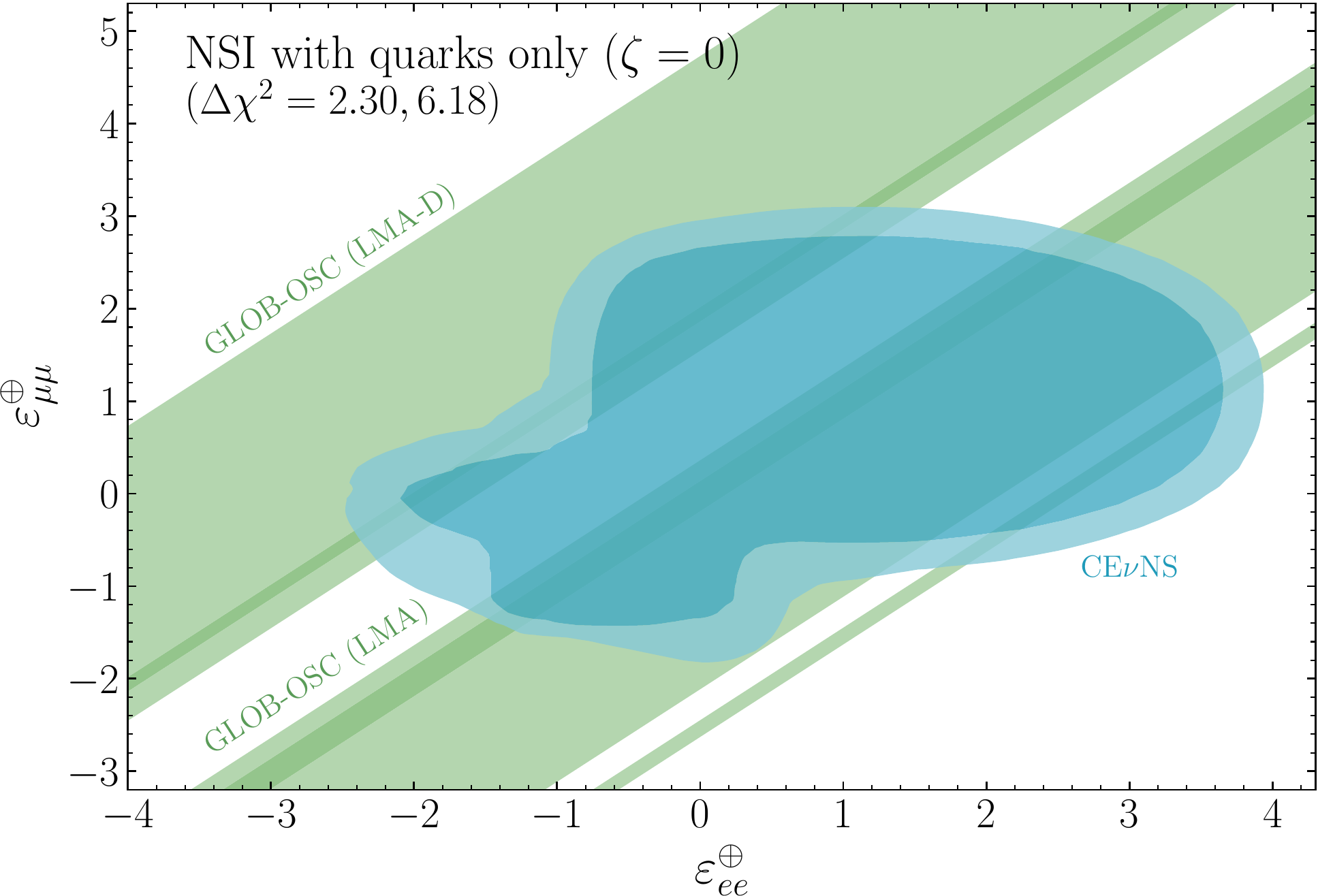}
  \caption{Allowed regions in the plane of $\Eps^\oplus_{ee}$ and
    $\Eps^\oplus_{\mu\mu}$ for vector NSI with quarks from the
    combinations of CE$\nu$NS data compared with the allowed regions
    from the global oscillation analysis which are the two diagonal
    shaded bands corresponding to the LMA and LMA-D solutions.  Both
    the green and blue regions are obtained after \emph{independently}
    marginalizing over all other relevant parameters: NSI couplings
    (including $\eta$) for the CE$\nu$NS region, and NSI couplings,
    $\eta$ and oscillation parameters for the <<GLOB-OSC>> regions.
    So the values of other NSI and $\eta$ in the blue and green
    regions are not forced to be the same.}
\label{fig:eemm}
\end{figure}

Let us start discussing the complementary sensitivity to vector-NSI
from the combined CE$\nu$NS results with that from present oscillation
data, for general models leading to NSI with quarks.  With this aim we
have first performed an analysis including only the effect of vector
NSI on the matter potential in the neutrino oscillation experiments.
The results of such analysis are given in the left column of
Table~\ref{tab:nsilblranges} in terms of the allowed ranges of the
effective NSI couplings to the Earth matter,
$\Eps^\oplus_{\alpha\beta}$, defined in Eq.~\eqref{eq:eps-earth0} (see
Sec.~\ref{sec:resulgen} for details).  Comparison with the constraints
from all available CE$\nu$NS data is illustrated in
Fig.~\ref{fig:eemm} where we plot the allowed regions in the plane
($\Eps^\oplus_{ee}$, $\Eps^\oplus_{\mu\mu}$).  The figure shows the
allowed regions obtained by the analysis of each of the two sets of
data \emph{independently}, after marginalization over the all other
(including off-diagonal) NSI parameters.\footnote{Technically the
regions for oscillations in Fig.~\ref{fig:eemm} are obtained by
performing each of the two analysis in terms of the 9 (out of the 10)
basic parameters listed in Eq.~\eqref{eq:10param} (fixing $\zeta=0$)
including the corresponding value of $Y_n$ to each data sample.
Therefore the output of each analysis is a $\chi^2$ function of the
basic 9 parameters, which is then marginalized with respect to all
parameters except for the two combinations shown in the figure.}  As
discussed in Sec.~\ref{sec:formCNUES}, even when considering NSI
coupling only to quarks, there is always a value of $\eta$ for which
the contribution of NSI to the CE$\nu$NS cross section cancels.
Consequently CE$\nu$NS data with a single nucleus does not lead to any
constraint on the $\Eps^\oplus_{\alpha\beta}$ parameters.  However,
the cancellation occurs at different values of $\eta$ for the three
nucleus considered (CsI, Ar, and Ge), and consequently, as seen in the
figure, the combination of CE$\nu$NS data with the three nuclear
targets does constrain the full space of effective
$\Eps^\oplus_{\alpha\beta}$, even after marginalization over $\eta$.
The constraints derived from the combined CE$\nu$NS data are
independent (and complementary) to those provided by the oscillation
analysis and, as seen in the figure, they are fully consistent.

The allowed ranges for vector NSI with up or down quarks
are compiled in Table~\ref{tab:nsiqranges} and show good qualitative
agreement with those of Refs.~\cite{Esteban:2018ppq, Coloma:2019mbs},
with the expected small deviations due to the differences in the
analysis, quoted CL, and included data.

\begin{sidewaystable}\centering
  \catcode`?=\active\def?{\hphantom{0}}
  \begin{tabular}{|l|c|c||c|c|}
    \hline
    & \multicolumn{2}{|c||}{Allowed ranges at 90\% CL (marginalized)}
    && Allowed ranges at 90\% CL (marginalized)
    \\
    \hline
    & \multicolumn{2}{|c||}{GLOB-OSC}
    && GLOB-OSC+CE$\nu$NS
    \\ \hline
    & LMA & $\text{LMA}\oplus\text{LMA-D}$
    && $\text{LMA} = \text{LMA}\oplus\text{LMA-D}$
    \\
    \hline
    \begin{tabular}{@{}c@{}}
      $\Eps_{ee}^{u,V} - \Eps_{\mu\mu}^{u,V}$ \\
      $\Eps_{\tau\tau}^{u,V} - \Eps_{\mu\mu}^{u,V}$
    \end{tabular}
     &
    \begin{tabular}{@{}c@{}}
      $[-0.063, +0.36]?$ \\
      $[-0.0053, +0.017]?$
    \end{tabular}
    &
    \begin{tabular}{@{}c@{}}
      $[-1.1, -0.79] \oplus [-0.063, +0.36]$ \\
      $[-0.021, +0.018]$
    \end{tabular}
    &
    \begin{tabular}{@{}c@{}}
      $\Eps_{ee}^{u,V}$ \\
      $\Eps_{\mu\mu}^{u,V}$ \\
      $\Eps_{\tau\tau}^{u,V} $
    \end{tabular}
    &
    \begin{tabular}{@{}c@{}}
      $[-0.038, +0.034] \oplus [+0.34, +0.42]$ \\
      $[-0.046, +0.031] \oplus [+0.35, +0.42]$ \\
      $[-0.046, +0.033] \oplus [+0.35, +0.42]$
    \end{tabular}
    \\
    $\Eps_{e\mu}^{u,V}$
    & $[-0.057, +0.013]$
    & $[-0.057, +0.061]$
    & $\Eps_{e\mu}^{u,V}$
    & $?[-0.044, +0.0049] $
    \\
    $\Eps_{e\tau}^{u,V}$
    & $[-0.076, +0.11]?$
    & $[-0.12, +0.11]$
    & $\Eps_{e\tau}^{d,V}$
    & $[-0.079, +0.11]?$
    \\
    $\Eps_{\mu\tau}^{u,V}$
    & $[-0.0077, +0.0042]$
    & $[-0.0077, +0.0083]$
    & $\Eps_{\mu\tau}^{u,V}$
    & $[-0.0064, 0.0053]$
    \\ \hline
    \begin{tabular}{@{}c@{}}
      $\Eps_{ee}^{d,V} - \Eps_{\mu\mu}^{d,V}$ \\
      $\Eps_{\tau\tau}^{d,V} - \Eps_{\mu\mu}^{d,V}$
    \end{tabular}
    &
    \begin{tabular}{@{}c@{}}
      $[-0.069, +0.38]?$ \\
      $[-0.0058, +0.018]?$
    \end{tabular}
    &
    \begin{tabular}{@{}c@{}}
      $?[-1.3, -0.91] \oplus [-0.072, +0.38]?$ \\
      $[-0.029, +0.019]$
    \end{tabular}
    &
    \begin{tabular}{@{}l@{}}
      $\Eps_{ee}^{d,V}$ \\
      $\Eps_{\mu\mu}^{d,V}$ \\
      $\Eps_{\tau\tau}^{d,V} $
    \end{tabular}
    &
    \begin{tabular}{@{}r@{}}
      $[-0.036, +0.031] \oplus [+0.30, +0.39]$ \\
      $[-0.040, +0.038] \oplus [+0.31, +0.39]$ \\
      $[-0.041, +0.043] \oplus [+0.31, +0.39]$
    \end{tabular}
    \\
    $\Eps_{e\mu}^{d,V}$
    & $[-0.058, +0.014]$
    & $[-0.058, +0.098]$
    & $\Eps_{e\mu}^{d,V}$
    & $?[-0.054, +0.0045]$
    \\
    $\Eps_{e\tau}^{d,V}$
    & $[-0.079, +0.11]?$
    & $[-0.16, +0.11]$
    & $\Eps_{e\tau}^{d,V}$
    & $[-0.051, +0.11]?$
    \\
    $\Eps_{\mu\tau}^{d,V}$
    & $[-0.0087, +0.0051]$
    & $[-0.0087, +0.015]?$
    & $\Eps_{\mu\tau}^{d,V}$
    & $[-0.0075, +0.0046]$
    \\
    \hline
  \end{tabular}
  \caption{90\% allowed ranges for the vector NSI couplings
    $\Eps_{\alpha\beta}^{u,V}$ and $\Eps_{\alpha\beta}^{d,V}$ as
    obtained from the global analysis of oscillation data (left
    columns, applicable to NSI induced by mediators with $\Mmed \ll
    5~\text{MeV}$) and also including data from CE$\nu$NS experiments
    (right columns, applicable to NSI induced by mediators with $\Mmed
    \gtrsim 50~\text{MeV}$).  The results are obtained after
    marginalizing over oscillation and the other matter potential
    parameters either within the LMA only ($\theta_{12} < 45^\circ$)
    and within both LMA ($\theta_{12} < 45^\circ$) and LMA-D
    ($\theta_{12} > 45^\circ$) subspaces respectively (this second
    case is denoted as $\text{LMA} \oplus \text{LMA-D}$).  Notice that
    once CE$\nu$NS data is included the two columns become identical,
    since for NSI couplings with $f=u,d$ the LMA-D solution is only
    allowed well above 90\% CL.}
  \label{tab:nsiqranges}
\end{sidewaystable}

Finally for completeness we have also performed a new dedicated
analysis including axial-vector NSI with up or down quarks.
Axial-vector NSI with quarks do not contribute to matter effects nor
to CE$\nu$NS.  They only enter the global analysis via their
modification of the NC event rate in the SNO experiment (see
Sec.~\ref{sec:sno-nc}), which is not able to constrain the NSI
coefficients if all of them are included simultaneously due to
possible cancellations between their respective contributions to the
NC rate.  Thus in this case we derive the bounds \emph{assuming that
only one NSI coupling is different from zero at a time}.  The
corresponding allowed ranges can be found in
Table~\ref{tab:nsiqaxialranges}.

As seen in the table, for all the coefficients, the allowed range is
composed of several disjoint intervals.  They correspond to values of
the NSI couplings for which the SNO-NC event rate is approximately the
SM prediction.  This can occur for either flavour diagonal or flavour
non-diagonal coeficients because solar neutrinos reach the Earth as
mass eigenstates, so the density matrix describing the neutrino state
at the detector is diagonal in the mass basis, but not in the flavor
basis.  The presence of non-vanishing off-diagonal
$\rho_{\alpha\neq\beta}$ elements is responsible for the sensitivity
to off-diagonal $\Eps_{\alpha\neq\beta}$ coefficients.  More
quantitatively, the ranges correspond to coefficients verifying:
\begin{equation}
  0 \simeq
  \Tr\Bigg[ \rho^\text{SNO}\, \bigg(\frac{G_A}{g_A}\bigg)^2 \Bigg] - 1
  = \left\{
  \begin{array}{l}
    \rho_{\alpha\alpha}^\text{SNO} \big[ \big(\Eps^{q,A}_{\alpha\alpha}\big)^2
      \pm 2\,\Eps^{q,A}_{\alpha\alpha} \big] \,,
    \\[2mm]
    \big( \rho^\text{SNO}_{\alpha\alpha} + \rho^\text{SNO}_{\beta\beta} \big)\,
    \big( \Eps^{q,A}_{\alpha\neq\beta} \big)^2
    \pm 4\Re\!\big(\rho^\text{SNO}_{\alpha\neq\beta}\big)\,
    \Eps^{q,A}_{\alpha\neq\beta}
  \end{array}\right.
\end{equation}
where the upper line holds when the NSI coefficient included is
flavour diagonal, and the lower one when it is flavour-changing and
the $\pm$ sign correspond to $q=u$ and $q=d$ respectively.  Thus the
allowed range for flavour diagonal NSI is formed by two solutions
around $\Eps_{\alpha\alpha}^{q,A} = 0$ and $\Eps_{\alpha\alpha}^{q,A}
= -2$.  For flavour off-diagonal NSI it is formed by solutions around
$\Eps_{\alpha\neq\beta}^{q,A} = 0$ and $\Eps_{\alpha\neq\beta}^{q,A} =
\mp 4\Re(\rho^\text{SNO}_{\alpha\neq\beta}) \mathbin{\big/}
(\rho^\text{SNO}_{\alpha\alpha} + \rho^\text{SNO}_{\beta\beta})$.  For
$\Eps^{q,A}_{e\mu}$ (and similarly for $\Eps^{q,A}_{e\tau}$) this last
condition corresponds, in fact, to two distinct solutions, both around
$|\Eps^{q,A}_{e\mu}|\neq 0$ and two possible signs, due to the two
different signs of $\rho^\text{SNO}_{e\mu}$ for the two CP-conserving
values of $\delta_\text{CP} \in \{0, \pi\}$.  On the contrary,
$\rho^\text{SNO}_{\mu\tau}$ takes very similar values for
$\delta_\text{CP} \in \{0, \pi\}$, and consequently for
$\Eps^{q,A}_{\mu\tau}$ the two non-zero solutions closely overlap
around $\Eps^{u,A}_{\mu\tau}\sim 1.7$ ($\Eps^{d,A}_{\mu\tau}\sim
-1.7$).

\begin{table}\centering
  \catcode`?=\active\def?{\hphantom{0}}
  \renewcommand{\arraystretch}{1.2}
  \begin{tabular}{|c || c || c |}\hline
    & {Allowed ranges at 90\% CL (1-parameter)}    &
    \\
    \hline
    & GLOB-OSC &
    \\
    \hline
    $\Eps^{u,A}_{ee}$
    & $??[-2.1, -1.8] \oplus [-0.19, +0.13]$
    & $-\Eps^{d,A}_{ee}$
    \\
    $\Eps^{u,A}_{\mu\mu}$
    & $??[-2.2, -1.7] \oplus [-0.26, +0.18]$
    & $-\Eps^{d,A}_{\mu\mu}$
    \\
    $\Eps^{u,A}_{\tau\tau}$
    & $??[-2.1, -1.8] \oplus [-0.20, +0.15]$
    & $-\Eps^{d,A}_{\tau\tau}$
    \\
    $\Eps^{u,A}_{e\mu}$
    & $[-1.5, -1.2] \oplus [-0.16, +0.12] \oplus [+1.4, +1.7]$
    & $-\Eps^{d,A}_{e\mu}$
    \\
    $\Eps^{u,A}_{e\tau}$
    & $[-1.5, -1.3] \oplus [-0.13, +0.10] \oplus [+1.4, +1.7]$
    & $-\Eps^{d,A}_{e\tau}$
    \\
    $\Eps^{u,A}_{\mu\tau}$
    & $[-0.085, +0.11] \oplus [+1.6, +1.9]???$
    & $-\Eps^{d,A}_{\mu\tau}$
    \\
    \hline
  \end{tabular}
  \caption{90\% CL bounds (1 d.o.f., 2-sided) on the effective
    axial-vector NSI couplings with quarks.  The bounds are derived
    from the global analysis of oscillation data including the effect
    of NSI in the SNO NC cross section and \emph{assuming only one NSI
    coupling different from zero at a time}.  As explained in
    Sec.~\ref{sec:sim}, these bounds apply to models with $\Mmed
    \gtrsim 3~\text{MeV}$.}
    \label{tab:nsiqaxialranges}
\end{table}

\subsection{Constraints on NSI with quarks and electrons: effective NSI in the Earth}
\label{sec:resulgen}

Let us now discuss the most general case in which NSI with quarks and
electrons are considered, parametrized by the angles $\eta$ and
$\zeta$ introduced in Eqs.~\eqref{eq:xi-eta}
and~\eqref{eq:xi-eta-quark}.  We focus on vector NSI because in this
case the interplay between matter and scattering effects, and
therefore the dependence on the couplings to charged fermions
involved, is expected to play a most relevant role.

\begin{figure}\centering
  \includegraphics[width=\textwidth]{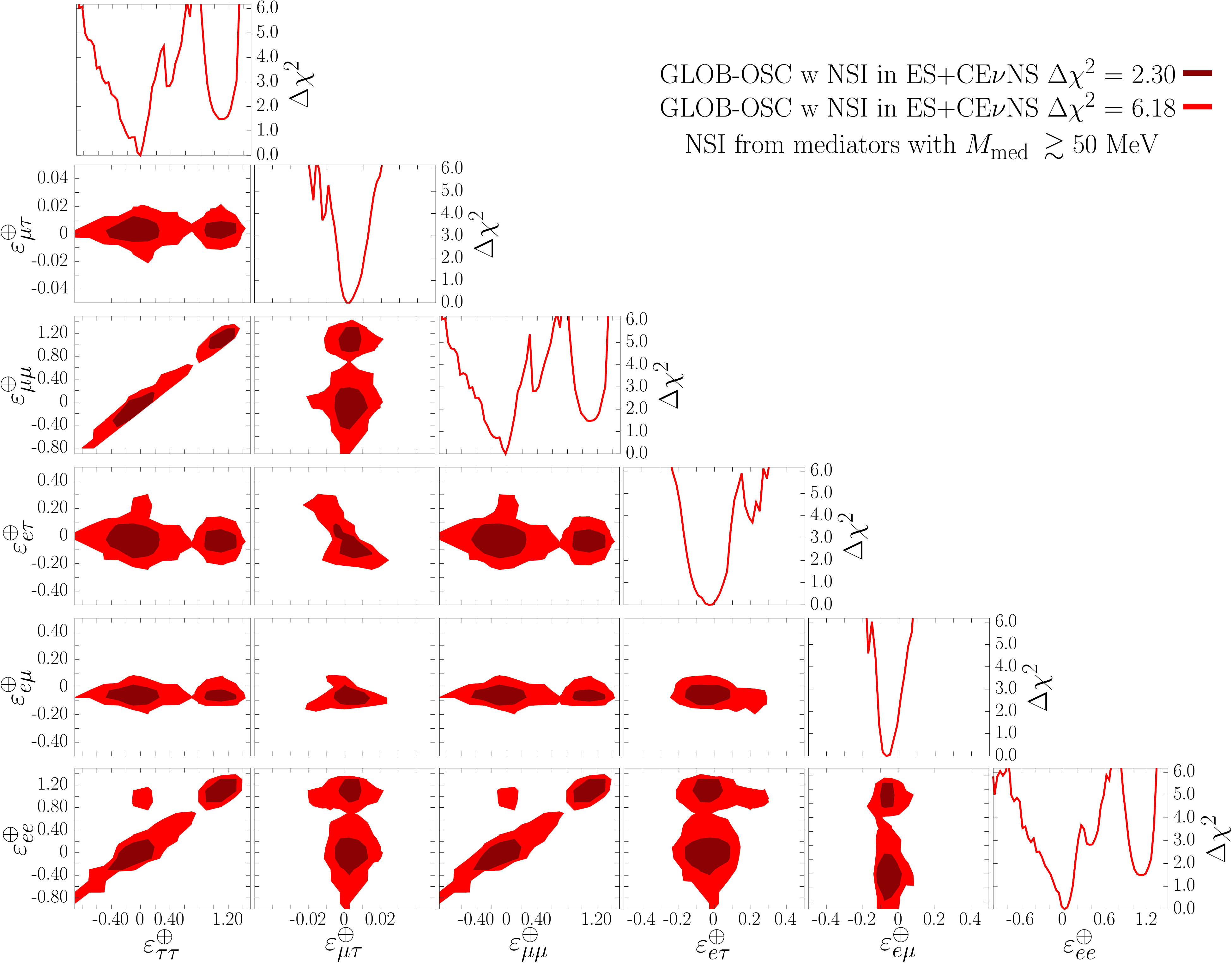}
  \caption{Constraints on the effective generalized NSI in the Earth
    matter (relevant for matter effects in LBL experiments) with
    arbitrary values of $\eta$ and $\zeta$.  Each panel shows a
    two-dimensional projection of the allowed multi-dimensional
    parameter space after minimization with respect to the undisplayed
    parameters.  The contours correspond to $1\sigma$ and $2\sigma$ (2
    d.o.f.).}
  \label{fig:earthtriang}
\end{figure}

In this framework, it is useful to quantify the results of our
analysis in terms of the effective NSI parameters which describe the
generalized Earth matter potential, which are in fact the relevant
quantities for the study of long-baseline and atmospheric oscillation
experiments.  The results are displayed in Fig.~\ref{fig:earthtriang}
and on the right column in Table~\ref{tab:nsilblranges} where we show
the allowed two-dimensional regions and one-dimensional ranges of the
effective NSI coefficients for the global analysis of oscillation and
CE$\nu$NS data, after marginalizing over all other parameters.
Therefore what we quantify in Fig.~\ref{fig:earthtriang} and the right
column in Table~\ref{tab:nsilblranges} is our present knowledge of the
matter potential for neutrino propagation in the Earth, for NSI
induced by mediators heavier than $\Mmed \gtrsim 50~\text{MeV}$ (as
discussed in Sec.~\ref{sec:sim}) and for \emph{any unknown value} of
$\eta$ and $\zeta$.  Technically this is obtained by marginalizing the
results of the global $\chi^2$ with respect to $\eta$ and $\zeta$ as
well, and the $\Delta\chi^2$ functions plotted in the figure are
defined with respect to the absolute minimum for any $\eta$ and
$\zeta$ which, lies close to $\eta \sim -45^\circ$ and $\zeta\sim
10^\circ$.

\begin{sidewaystable}\centering
  \newcommand{\Twin}[2]{\begin{tabular}{@{}c@{}}$#1$\\[-2mm]$\big(#2\big)$\end{tabular}}
  \renewcommand{\arraystretch}{1.2}
  \begin{tabular}{|@{}c@{}||@{}c@{}|}
    \hline
    \multicolumn{2}{|c|}{Allowed ranges at \Twin{90\%~\text{CL}}{99\%~\text{CL}} marginalized}
    \\
    \hline\hline
    GLOB-OSC w/o NSI in ES
    & GLOB-OSC w NSI in ES + CE$\nu$NS
    \\
    \hline
    \begin{tabular}{c||c}
      \\[-1.15mm]
      $\Eps^\oplus_{ee}-\Eps^\oplus_{\mu\mu}$
      & \Twin{[-3.1, -2.8] \oplus [-2.1, -1.88] \oplus [-0.15, +0.17]}{[-4.8, -1.6] \oplus [-0.40, +2.6]}
      \\[+6mm]
      $\Eps^\oplus_{\tau\tau}-\Eps^\oplus_{\mu\mu}$
      & \Twin{[-0.0215, +0.0122]}{[-0.075, +0.080]}
      \\[+6mm]
      $\Eps^\oplus_{e\mu}$
      & \Twin{[-0.11, -0.021] \oplus[+0.045, +0.135]}{[-0.32, +0.40]}
      \\[+3mm]
      $\Eps^\oplus_{\mu\tau}$
      & \Twin{[-0.22, +0.088]}{[-0.49, +0.45]}
      \\[+3mm]
      $\Eps^\oplus_{\mu\tau}$
      & \Twin{[-0.0063, +0.013]}{[-0.043, +0.039]}
    \end{tabular}
    &
    \begin{tabular}{c||c}
      $\Eps^\oplus_{ee}$
      & \Twin{[-0.19, +0.20] \oplus [+0.95, +1.3]}{[-0.23, +0.25] \oplus [+0.81, +1.3]}
      \\[+3mm]
      $\Eps^\oplus_{\mu\mu}$
      & \Twin{[-0.43, +0.14]\oplus [+0.91, +1.3]}{[-0.29, +0.20] \oplus [+0.83, +1.4]}
      \\[+3mm]
      $\Eps^\oplus_{\tau\tau}$
      & \Twin{[-0.43, +0.14]\oplus [+0.91, +1.3]}{[-0.29,  +0.20]\oplus [+0.83, +1.4]}
      \\[+3mm]
      $\Eps_{e\mu}^\oplus$
      & \Twin{[-0.12, +0.011]}{[-0.18, +0.08]}
      \\[+3mm]
      $\Eps_{e\tau}^\oplus$
      & \Twin{[-0.16, +0.083]}{[-0.25, +0.33]}
      \\[+3mm]
      $\Eps_{\mu\tau}^\oplus$
      & \Twin{[-0.0047, +0.012]}{[-0.020, +0.021]}
    \end{tabular}
    \\
    \hline
  \end{tabular}
  \caption{90\% and 99\% CL bounds (1 d.o.f., 2-sided) on the
    effective NSI parameters relevant for matter effects in LBL
    experiments with arbitrary values of $\eta$ and $\zeta$, obtained
    after marginalizing over all other NSI and oscillation parameters.
    The bounds on the left (right) column are applicable to NSI
    induced by mediators with masses $\Mmed \ll 500~\text{keV}$
    ($\Mmed \gtrsim 50~\text{MeV}$).}
  \label{tab:nsilblranges}
\end{sidewaystable}

From these results we see that the allowed ranges for the diagonal
$\Eps^\oplus_{\alpha\alpha}$ parameters are composed of two disjoint
regions.  However let us stress that they both correspond to the LMA
solution since neither of them falls within the LMA-D region
(which requires $\Eps^\oplus_{ee} - \Eps^\oplus_{\mu\mu} = -2$).  In
fact, the LMA-D solution is ruled out beyond the CL shown in
Fig.~\ref{fig:earthtriang} and, as will be discussed in
Sec.~\ref{sec:resulLMAD} in more detail, the combination of
oscillation data with CE$\nu$NS results for different nuclear targets
is required to reach this sensitivity.

Conversely, for mediators with masses $\Mmed \ll 500~\text{keV}$,
effects on ES and CE$\nu$NS experiments would be suppressed even if
NSI involve couplings to both quarks and electrons, and the only
effect of NSI will be the modification of the matter potential in
neutrino oscillations.  The same holds for NSI with quarks only
(\textit{i.e.}, for $\zeta=0$) induced by mediators with masses $\Mmed
\ll 10~\text{MeV}$.  Since in the matter potential the effects for
protons or electrons are indistinguishable, the allowed ranges of the
effective $\Eps^\oplus_{\alpha\beta}$ are the same in both scenarios,
which are given on the left column in Table~\ref{tab:nsilblranges}
(and included in Fig.~\ref{fig:eemm}).  As discused in
Sec.~\ref{sec:formOSC}, in this case the analysis can only constrain
the differences between flavour-diagonal NSI coefficients.
Furthermore, the allowed range of $\Eps^\oplus_{ee} -
\Eps^\oplus_{\mu\mu}$ contains a disjoint interval around $-2$ (which
at 90\% CL further splits into two sub-intervals as seen in the table)
corresponding to the LMA-D solution, which is well allowed in these
scenarios.  We will discuss this in more detail in
Sec.~\ref{sec:resulLMAD}.

\begin{figure}\centering
  \includegraphics[width=0.5\textwidth]{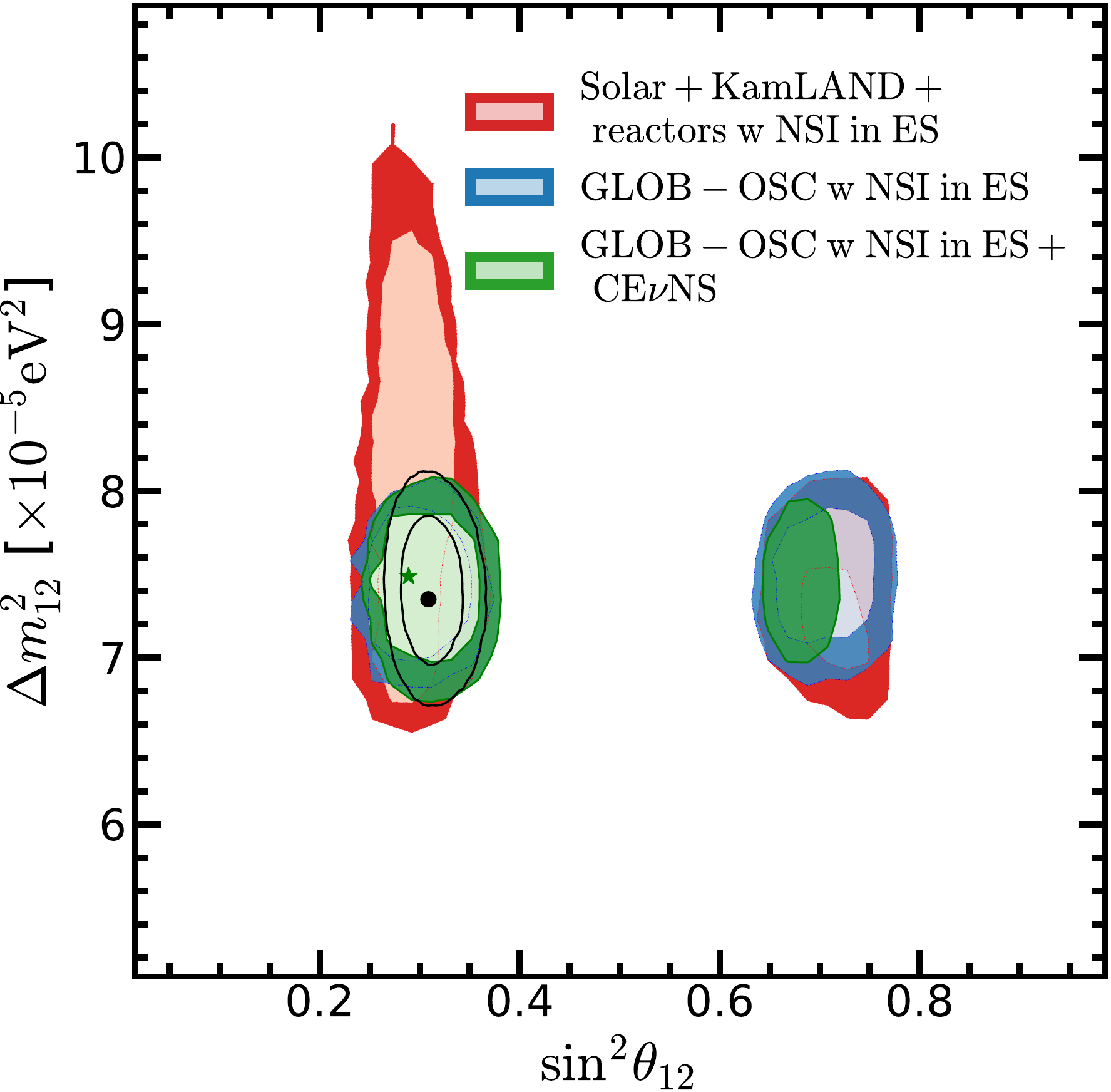}
  \caption{Two-dimensional projections of the allowed regions (at 90\%
    and $3\sigma$ CL) onto $\Dmq_{12}$ and $\theta_{12}$ parameters,
    after marginalizing over all other oscillation parameters and over
    NSI couplings to quarks and electrons.  Red regions correspond to
    the analysis of KamLAND and Solar data; blue regions include all
    oscillation data; and green regions include all oscillation and
    CE$\nu$NS data.  In all cases, NSI effects on ES are fully
    accounted for in the fit.  For comparison, the empty contours
    (solid black lines) show the corresponding regions for the global
    oscillation analysis without NSI.}
\label{fig:oscil}
\end{figure}

We finish this section by discussing the impact of NSI in this general
framework (where we allow couplings to quarks and electrons
simultaneously) on the determination of the oscillation parameters in
the solar sector.  This is shown in Fig.~\ref{fig:oscil}, where we see
that the determination of the oscillation parameters within the LMA
region, which is the region favored by the fit, is rather robust even
after the inclusion of NSI couplings to both quarks and electrons.
Comparing the blue and red regions in the figure we see that the
inclusion of data from atmospheric and LBL experiments is important to
reach such robustness.  This had been previously shown in
Refs.~\cite{Gonzalez-Garcia:2013usa, Esteban:2018ppq} for NSI with
quarks only; here we conclude that the same conclusions hold also in
presence of NSI with electrons, as long as their impact on ES is
accounted for in the fit.  Also, the allowed LMA regions are not very
much affected by the addition of the CE$\nu$NS data and it is very
close to that of the oscillation analysis without NSI.  As also shown
in the figure, the LMA-D is only allowed at 97\% CL or above (for
2~d.o.f.).  The current status of the LMA-D region is discussed in
more detail in the next section.

\subsection{Present status of the LMA-D solution}
\label{sec:resulLMAD}

\begin{figure}\centering
  \includegraphics[width=\textwidth]{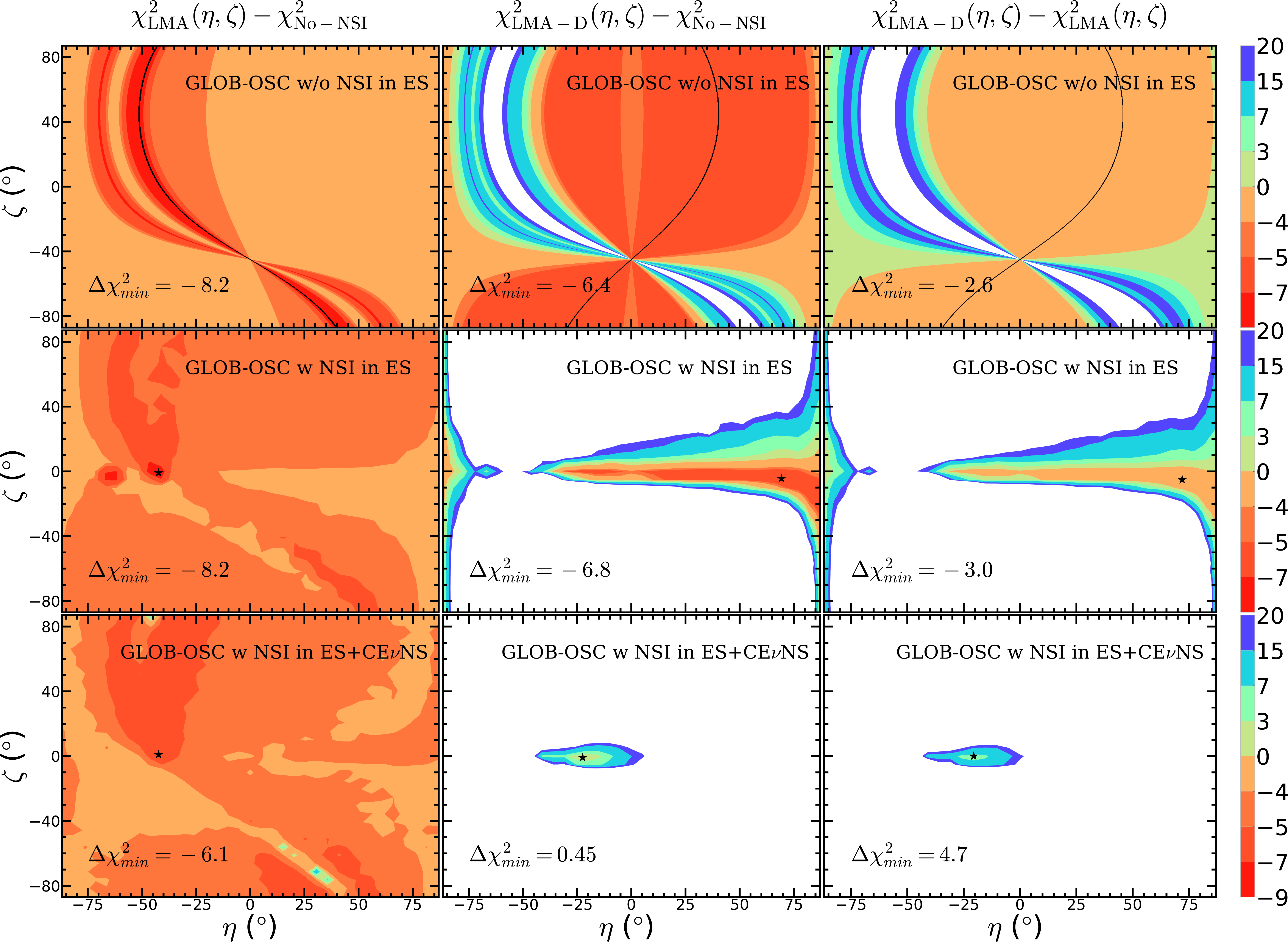}
  \caption{Isocontours of $\chi^2_\text{LMA}(\eta,\zeta) -
    \chi^2_\text{no-NSI}$, $\chi^2_\text{LMA-D}(\eta,\zeta) -
    \chi^2_\text{no-NSI}$ and $\chi^2_\text{LMA-D}(\eta,\zeta) -
    \chi^2_\text{LMA-D}(\eta,\zeta)$ of the global analysis of
    oscillation data without including NSI in the ES cross sections at
    Borexino, SNO, and SK (upper panels), and including the NSI in the
    ES cross sections at Borexino, SNO, and SK (middle panels).  The
    lower panels shows of the result adding the data from CE$\nu$NS
    experiments.  We show projections in the plane of angles ($\zeta$,
    $\eta$) (after marginalization of all other parameters) which
    parametrize the relative strength of the NSI couplings to
    up-quarks, down-quark, and electrons.  The levels corresponding to
    the different colours are given on the color bar on the right.
    Contours beyond 20 are white.  In each panel the best-fit point is
    marked with a star (middle and bottom rows) or by a solid black
    line (upper row).  Results in the upper row are applicable to NSI
    induced by mediators with $\Mmed \ll 500~\text{keV}$; in the
    middle row, for models with $\Mmed \gtrsim 10~\text{MeV}$; and in
    the lower row, for models with $\Mmed \gtrsim 50~\text{MeV}$, see
    Sec.~\ref{sec:sim} for details.}
  \label{fig:2D}
\end{figure}

In this section, we discuss in more detail the present status of the
LMA-D region in light of all available data, for models leading to NSI
couplings to quarks and electrons simultaneously.  We start by
exploring the dependence of the presence of the LMA-D solution on the
specific combination of couplings to the charged fermions considered.
In order to do so it is convenient to introduce the functions
$\chi^2_\text{LMA}(\eta,\zeta)$ and $\chi^2_\text{LMA-D}(\eta,\zeta)$
which are obtained by marginalizing the $\chi^2$ for a given value of
$\eta$ and $\zeta$ over both the oscillation and the matter potential
parameters within the regions $\theta_{12} < 45^\circ$ and
$\theta_{12} > 45^\circ$, respectively.  With this, in the left
(central) panels of Fig.~\ref{fig:2D} we plot isocontours of the
differences $\chi^2_\text{LMA}(\eta,\zeta) - \chi^2_\text{no-NSI}$
($\chi^2_\text{LMA-D}(\eta,\zeta) - \chi^2_\text{no-NSI}$) where
$\chi^2_\text{no-NSI}$ is the minimum $\chi^2$ for standard $3\nu$
oscillations (\textit{i.e.}, without NSI).  In the right panels we
plot $\chi^2_\text{LMA-D}(\eta,\zeta) - \chi^2_\text{LMA}(\eta,\zeta)$
which quantifies the relative quality of the LMA and LMA-D solutions.
In each row in this figure we include only a subset of the data as
indicated by the labels.

The upper panels of Fig.~\ref{fig:2D} show the results when only
oscillation data is analyzed accounting for the effects of NSI on the
matter potential, but without including the effect of NSI in the ES
cross sections in Borexino, SNO, and SK.  As outlined in
Sec.~\ref{sec:sim}, these results would therefore apply to NSI models
with very light mediators, $\Mmed \ll 500~\text{keV}$.  In this
scenario (as shown in Sec.~\ref{sec:formOSC}) only the combination of
NSI couplings to electrons, protons and neutrons, parametrized by the
effective angle $\eta^\prime$ in Eq.~\eqref{eq:epx-etapr}, is
relevant.  Therefore the $\Delta\chi^2$ isocontours are curves along
$\tan\eta^\prime = \tan\eta \mathbin{/} (\cos\zeta + \sin\zeta) =
\text{constant}$.  From the upper left panel we see that for most of
$(\eta,\zeta)$ values, the inclusion of NSI leads only to a mild
improvement of the global fit to oscillation data
($\chi^2_\text{LMA}(\eta,\zeta) - \chi^2_\text{no-NSI} > -4$).  As
discussed in Ref.~\cite{Esteban:2018ppq} (Addendum), with the updated
SK4 solar data the determination of $\Dmq_{21}$ in solar and in
KamLAND experiments are fully compatible at $\lesssim 2\sigma$ level.
The inclusion of NSI only leads to an overall better fit at a CL above
2$\sigma$ (but still not statistically significant in any of the
different data samples) for $(\zeta,\eta)$ along the darker band.  In
particular the best fit lies along $\tan\eta^\prime \sim -1 \big/
Y_n^\oplus \approx -0.95$ ($\eta^\prime \approx -43.6^\circ$) for
which NSI effects in the Earth matter cancel, so there is no
constraint from Atmospheric and LBL experiments on the NSI which can
lead to that slightly better fit to Solar + KamLAND data within the
LMA solution.  The central and right panels show the status of the
LMA-D solution in this scenario.  We find that LMA-D provides a good
solution in most of the $(\eta,\zeta)$ plane.  The LMA-D is only very
disfavoured for $(\zeta, \eta)$ along the band with $\-2.75\lesssim
\tan\eta^\prime \lesssim -1.75$.  For these values the NSI
contribution to the matter potential in the Sun cancels in some point
inside the neutrino production area and therefore the degeneracy
between NSI and the octant of $\theta_{12}$ cannot be realized.

The impact of including NSI in the ES cross section in Borexino, SNO,
and SK can be seen in the middle row panels in Fig.~\ref{fig:2D}.
Since these panels include the effect of NSI in ES, they apply to NSI
models with mediator masses $10~\text{MeV}\lesssim \Mmed \lesssim
50~\text{MeV}$.  The main effect is that the $\zeta$ values for which
the LMA-D solution is allowed become very restricted (in fact the best
fit point for all panels are always close to $\zeta=0$) as long as
$\eta$ does not approach $90^\circ$, and consequently the dependence
of $\zeta$ is heavily suppressed.  Thus, once the effect of NSI in the
ES cross section is included, the results are not very different from
those obtained for $\zeta=0$ (no coupling to electrons) in
Refs.~\cite{Esteban:2018ppq, Coloma:2019mbs}.  The middle and right
panels in this row also illustrate how, as long as only oscillation
data is included, the LMA-D solution is still allowed with a CL
comparable and even slightly better than LMA.

The lower panels in Fig.~\ref{fig:2D} include the combination of all
data available and are therefore applicable to NSI with mediators
above $\Mmed \gtrsim 50~\text{MeV}$.  When comparing the LMA (LMA-D)
to the SM hypothesis (\textit{i.e.,} no NSI) we find that the global
minimum of the fit is better (comparable) to that obtained in absence
of NSI, as shown in the middle and left panels.  However, we also see
that the inclusion of the CE$\nu$NS data in the analysis severely
constrains the LMA-D solution.  Quantitatively we find that LMA-D
becomes disfavoured with respect to LMA with $\Delta\chi^2>4.7$ for
any value of $(\eta,\zeta)$ (right panel) and it is only
allowed\footnote{Notice than this is different than what we show in
Fig.~\ref{fig:2D}, where marginalization over $\eta,\zeta$ has been
performed.} below $\Delta\chi^2=9$ for very specific combinations of
couplings to quarks and electrons, $-2.5^\circ\leq\zeta\leq 1.5^\circ$
and $-29^\circ\leq\eta\leq -13^\circ$.

\begin{figure}\centering
  \includegraphics[width=0.5\textwidth]{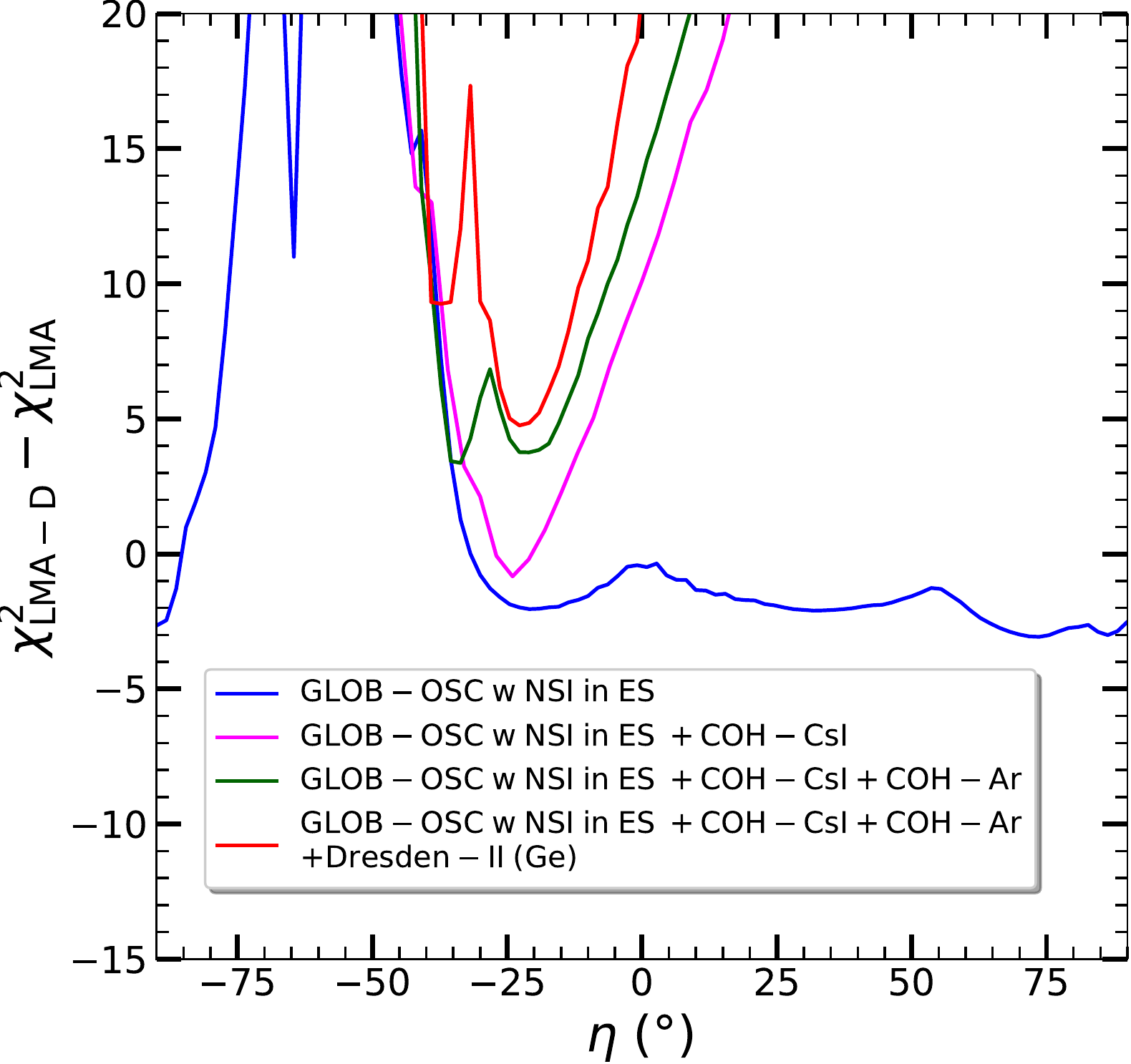}
  \caption{Dependence of $\chi^2_\text{LMA-D}(\eta,\zeta) -
    \chi^2_\text{LMA}(\eta,\zeta)$ on $\eta$ after marginalizing over
    $\zeta$ for different combination of experiments as labeled in the
    figure.}
  \label{fig:1Deta}
\end{figure}

To further illustrate the role of the different experiments in this
conclusion we show in Fig.~\ref{fig:1Deta} the projection of
$\chi^2_\text{LMA-D}(\eta,\zeta)-\chi^2_\text{LMA}(\eta,\zeta)$ on
$\eta$ after marginalizing over $\zeta$ (which, as discussed above, it
is effectively not very different from fixing $\zeta=0$).  The figure
illustrates the complementarity of the CE$\nu$NS data with different
targets.  As discussed in Sec.~\ref{sec:formCNUES}, the effects of NSI
in CE$\nu$NS on a given target, characterized by a value of $Y_n$,
cancel for $\eta^{\prime\prime} = \arctan(-1 / Y_n)$ (with
$\tan\eta^{\prime\prime} = \tan\eta / \cos\zeta \simeq \tan\eta$).
This corresponds to $\eta\simeq -35.4^\circ$ ($Y_n^\text{CsI} \approx
1.407$), $-39.3^\circ$ ($Y_n^\text{Ar} \approx 1.222$), and
$-38.4^\circ$ ($Y_n^\text{Ge} \approx 1.263$) for CsI, Ar, and Ge
respectively.  Thus, as seen in the figure, the combination of the
constraints from CE$\nu$NS with the different targets is important to
disfavour the LMA-D solution.

\section{Summary}
\label{sec:summary}

In this work we have presented an updated analysis of neutrino
oscillation data and in combination with the results of CE$\nu$NS on a
variety of targets with the aim of establishing the allowed size and
flavour structure of CP-conserving NC NSI which affect either the
evolution of neutrinos in a matter background and/or their detection
cross section.  We have included in our fit all the latest solar,
atmospheric, reactor and accelerator data used for the standard 3$\nu$
oscillation analysis in NuFIT-5.2 with the only exception of T2K and
NO$\nu$A appearance data whose hints in of CP violation cannot be
accommodated within the CP-conserving approximation assumed in this
work.  When combining with CE$\nu$NS we include the results from
COHERENT data on CSI and Ar detectors, together with the results
Dresden-II reactor experiment with a Ge detector.

The study extends previous works by considering either vector or axial
NSI with an arbitrary ratio of couplings to up-quarks, down-quarks,
and electrons (parametrized by two angles, $\eta$ and $\zeta$).  In
the oscillation analysis including vector NSI involving interactions
with electrons ($\zeta\neq 0$), which could affect both the
propagation and ES detection cross section in Borexino, SNO, and SK,
we have consistently accounted for the interplay of flavour
transitions in both effects by employing the density matrix formalism
(which we had previously proposed).  Furthermore we have considered
two scenarios: (a) including the NSI only in the matter effects
(characteristic of models where NSI are generated by mediators much
lighter than $\mathcal{O}(500~\text{keV})$), and (b) including the NSI
both in propagation, CE$\nu$NS, and ES scattering (characteristic of
models where NSI are generated by mediators heavier than
$\mathcal{O}(50~\text{MeV})$).  Our results show that:
\begin{itemize}
\item From the methodological point of view, the validity of the
  adiabaticity assumption for the neutrino propagation in the Sun in
  the presence of NSI (and of NP, in general) is not a given.  But we
  conclude that using the adiabatic approximation leads to the correct
  allowed ranges of NSI coefficients in all cases studied, \emph{as
  long as one removes from the explored space of NSI couplings those
  points for which adiabaticity in the Sun is violated}.

\item The global oscillation analysis of vector NSI which enter only
  via the matter potential leads to the bounds on the five
  combinations $\Eps_{ee}^{f,V} - \Eps_{\mu\mu}^{f,V}$ and
  $\Eps_{\tau\tau}^{f,V} - \Eps_{\mu\mu}^{f,V}$ and
  $\Eps^{f,V}_{\alpha\neq\beta}$ shown in Fig.~\ref{fig:etriangVwoES}
  and Table~\ref{tab:nsieVwoES} for interactions with electrons, and
  in the first column in Table~\ref{tab:nsiqranges} for interactions
  with up and down quark.  The 90\% bounds range from $\sim 0.5$\% to
  $\sim 30$\% for $\Eps_{\tau\tau}^{f,V} - \Eps_{\mu\mu}^{f,V}$ and
  $\Eps^{f,V}_{\alpha\neq\beta}$, while $\Eps_{ee}^{f,V} -
  \Eps_{\mu\mu}^{f,V}$ presents two (quasi)degenerate allowed ranges
  around $0$ and $\sim -2$ ($\sim -1$) for NSI with electrons (quarks)
  which correspond to the well-known LMA and LMA-D solutions.

\item The LMA-D solution is realized for generic vector NSI couplings
  to electrons and quarks as long as only considering propagation
  effects (see first row in Fig.~\ref{fig:2D}) unless the couplings
  come in a ratio that cancels their contribution to the solar matter
  potential ($\-2.75\lesssim \tan\eta \mathbin{/} (\cos\zeta +
  \sin\zeta) \lesssim -1.75$).

\item Including the effect of the vector NSI in the ES scattering
  cross sections in Borexino, SK and SNO, lifts the degeneracy and
  disfavours the LMA-D solution unless the NSI coupling to the
  electrons is suppressed compared to the coupling to quarks (see
  second row in Fig.~\ref{fig:2D}).  But it still provides a good fit
  to the data for a wide range of ratios of the couplings to up and
  down quarks, and it is allowed at 3$\sigma$ for $\eta < -70^\circ$
  or $\eta > -40^\circ$ (see Fig.~\ref{fig:1Deta}).

\item For vector NSI coupling to electrons only, the inclusion of NSI
  in the ES scattering cross sections eliminates the LMA-D solution
  and allows for independent determination of the three
  $\Eps_{\alpha\alpha}^{f,V}$.  But within the LMA solution, our
  comparative results show that it is the presence of the NSI in the
  matter potential which drives the strong constraints.

\item For vector NSI coupling mostly to quarks induced by mediators
  with $\Mmed \gtrsim 50~\text{MeV}$, the combination of the
  oscillation results with the data from CE$\nu$NS disfavours the
  LMA-D solution beyond $\sim 2\sigma$ for any value of $\eta$ and it
  is only allowed below $\Delta\chi^2=9$ for $-29^\circ \leq \eta \leq
  -13$.  (see Figs.~\ref{fig:2D} and~\ref{fig:1Deta}).  We find that
  the combination of CE$\nu$NS with different nuclear targets is
  important to disfavour the LMA-D solution at that level.

\item The determination of the oscillation parameters within the LMA
  region is rather robust even after the inclusion of NSI couplings to
  both quarks and electrons as large as allowed by the global
  oscillation data analysis itself (see Fig.~\ref{fig:oscil}).

\item Axial-vector NSI only enter in the analysis via their effects in
  the interaction of solar neutrinos in Borexino, SNO, and SK when
  coupling to electrons, or in the NC rate in SNO when coupling to
  quarks.  For interactions with electrons their bounds are notably
  weaker than those on vector NSI (see Fig.~\ref{fig:etriangA} and
  Table~\ref{tab:nsiewES}).  For interactions with quarks they cannot
  be all independently bounded (see Table~\ref{tab:nsiqaxialranges}
  for one parameter bounds).
\end{itemize}
Moreover, to quantify the possible implications that generic models
leading to NSI may have for future Earth-based facilities, we have
projected the results of our analysis in terms of the effective NSI
parameters which describe the generalized matter potential in the
Earth, relevant for the study of atmospheric and long-baseline
neutrino oscillation experiments (see Fig.~\ref{fig:earthtriang} and
Table~\ref{tab:nsilblranges}).

To summarize, this work provides the most general study of NSI
involving quarks and electrons up to date using neutrino oscillation
data and measurements of CE$\nu$NS using neutrinos from spallation and
reactor sources.  While the scope of this work is restricted to
neutrino facilities, dark matter direct detection experiments are
rapidly approaching the so-called neutrino floor.  If affected by NSI,
significant deviations on the expected results may be
observable~\cite{Cerdeno:2016sfi, Boehm:2018sux,
  Gonzalez-Garcia:2018dep}.  While present experiments are not
competitive yet in this regard, the upcoming generation of experiments
will soon provide interesting limits on NSI and may offer additional
synergies with neutrino oscillation and CE$\nu$NS
data~\cite{Amaral:2023tbs}.

\acknowledgments

We are grateful to E.~Fernandez-Martinez for useful discussions on the
interference effects between neutrino oscillations and interactions in
the presence of NSI.
This project is funded by USA-NSF grant PHY-1915093 and by the
European Union through the Horizon 2020 research and innovation
program (Marie Sk{\l}odowska-Curie grant agreement 860881-HIDDeN) and
the Horizon Europe programme (Marie Sk{\l}odowska-Curie Staff Exchange
grant agreement 101086085-ASYMMETRY).  It also receives support from
grants PID2019-\allowbreak 105614GB-\allowbreak C21,
PID2019-\allowbreak 108892RB-\allowbreak I00, PID2019-\allowbreak
110058GB-\allowbreak C21, PID2020-\allowbreak 113644GB-\allowbreak
I00, ``Unit of Excellence Maria de Maeztu 2020-2023'' award to the
ICC-UB CEX2019-000918-M, and grant IFT ``Centro de Excelencia Severo
Ochoa'' CEX2020-001007-S funded by MCIN/AEI/\allowbreak
10.13039/\allowbreak 501100011033, as well as from grant 2021-SGR-249
(Generalitat de Catalunya) and from the ``Generalitat Valenciana''
grant PROMETEO/2019/087.  PC is supported by grant RYC2018-024240-I
funded by MCIN/AEI/\allowbreak 10.13039/\allowbreak 501100011033 and
by ``ESF Investing in your future''.  SU acknowledges support from
Generalitat Valenciana through the plan GenT program
(CIDEGENT/2018/019) and from the Spanish MINECO under Grant
FPA2017-85985-P.  We also acknowledge use of the IFT and IFIC
computing facilities (Hydra and SOM clusters).

\bibliographystyle{JHEPmod}
\bibliography{references}

\end{document}